# Lagrangian formalism in the theory of relativistic vector fields

Sergey G. Fedosin

22 Sviazeva str., apt. 79, Perm, Perm Krai, 614088, Russia

E-mail: fedosin@hotmail.com

**Abstract:** The Lagrangian formalism is used to derive covariant equations that are suitable for use in continuously distributed matter in curved spacetime. Special attention is given to theoretical representation, in which the Lagrangian and its derivatives are directly involved. The obtained results, including equation for metric, equation of motion, equations for fields, are applied to purely vector fields. As a consequence, formulas are determined for calculating the basic quantities necessary to describe physical systems. In this case, not only the pressure field and the acceleration field are taken into account, but also the electromagnetic and gravitational fields outside the matter, which contribute to the four-momentum and to the four-dimensional angular momentum pseudotensor of each system. It is shown that the canonical representation of the angular momentum pseudotensor is its representation with covariant indices. The radius-vector of the center of momentum of a physical system is determined in covariant form.

## 1. Introduction

In field theory, the principle of least action is a basic principle that allows to find field equations and equations of motion of particles in the given fields. If the action is denoted by $S$, and Lagrangian density, calculated as volumetric density of the Lagrangian, is denoted by $\mathcal{L}$, then the following relation is valid for the action [1]:

$$S = \int_{t_1}^{t_2} L \, dt = \int_{t_1}^{t_2} \left( \int_V \mathcal{L} \, dV \right) dt . \qquad (1)$$

The meaning of (1) is that when a physical system moves from time point $t_1$ to time point $t_2$, for each Lagrangian $L$ there is such a trajectory of motion, at which the value $S$ becomes extreme. In this case, the indicated trajectory is the most likely trajectory. To determine the extremum, it is necessary to equate the action variation to zero, that is, to assume that $\delta S = 0$. Then it follows from (1) that $\delta L = 0$, which, with a specified expression for $\mathcal{L}$, defines the field equations and the equation of the system's motion.

In the simplest case, we can assume that the field superposition principle holds true in the sense that each field in the Lagrangian equation is present independently of the other fields. Then we can apply the Lagrangian to the terms responsible for some fields, and use the variation in these terms to derive the equation for this field.

The spacetime properties in the physical system are given by the metric tensor $g_{\mu\nu}$, with the help of which the indices are raised and lowered in four-dimensional vectors and tensors. Using metric tensor derivatives with respect to four-dimensional coordinates, the Christoffel symbols, the spacetime curvature tensor and the scalar curvature are calculated, that is, those quantities allow us to take into account the difference between curved spacetime and flat Minkowski spacetime.

In standard approach, the terms responsible for the energy densities of particles and fields of a physical system are introduced into the Lagrangian density $\mathcal{L}$. All these terms must be covariant invariants and scalar functions, just like the action $S$ itself, which has the dimension of the product of energy by time. As a rule, the Lagrangian contains the energies of particles in the fields, which are found using the product of four-potential of corresponding field and mass four-current $J^\mu$, while in the case of electromagnetic field the charge four-current $j^\mu$ is used. In addition, the Lagrangian contains the terms, defining energy density of the fields and expressed in terms of tensor invariants of these fields. The scalar curvature is an invariant of curvature tensor, and in the Lagrangian it defines the term proportional to the system's energy density and arises by considering the system's metric. Finally, the cosmological constant in the Lagrangian is the only term, for which the system's energy can be uniquely gauged. Indeed, the energy as a scalar physical quantity is determined with an accuracy of up to a constant, so that uniqueness in determining the state of the system is possible only because of taking into account the energy gauging.

$S$ in (1) is varied for each term, which is included in the Lagrangian density, separately for each variable quantity $q$, this can be represented, for example, by the field's four-potentials, the four-currents or the metric tensor. Then all the terms, containing the variation in any variable quantity, should be summed, and the sum should be equated to zero. As a result, the action variation with respect to the four-currents leads to the equation of motion, the action variation with respect to the fields' four-potentials gives the equations for the fields, and the variation with respect to the metric tensor gives the equation for the metric. This is the practical approach in the Lagrangian formalism.

On the other hand, there is a theoretical approach, in which the procedure for determining the extremum of the action in (1) gives the Euler–Lagrange equations, which also lead to the equations of motion of particles and fields. If the Lagrangian density depends on a certain variable quantity $q$ and on its four-dimensional derivative $\partial_\nu q = \dfrac{\partial q}{\partial x^\nu}$ with respect to coordinates and time, such that $\mathcal{L} = \mathcal{L}(q, \partial_\nu q)$, then the corresponding classical Euler–Lagrange equation can be written as follows [2]:

$$\frac{\partial}{\partial x^\nu}\left(\frac{\partial \mathcal{L}}{\partial(\partial_\nu q)}\right) = \frac{\partial \mathcal{L}}{\partial q}. \qquad (2)$$

Our goal in this paper is to specify relations (1) and (2) so that they are suitable for curved spacetime in an arbitrary reference frame for continuously distributed matter. Next, we will present the Euler–Lagrange equations in a covariant form and compare the resulting equations with the equations obtained by ordinary variation of the action function. In addition, the covariant formulas for the energy, momentum, four-momentum, angular momentum and four-dimensional angular momentum pseudotensor are presented.

The necessity for this article is caused by the fact that the mainstream of literature on the Lagrangian formalism and the related Hamiltonian formalism is devoted to individual particles and bodies [3-5], while it is often necessary to carry out calculations for the continuous medium. In this case, the Lagrangian formalism should be applied to the integrals taken over the matter's volume, which significantly complicates the situation in curved spacetime. This probably explains the apparent lack of related research, as well as incomplete coverage of the problems under consideration. The same is also indicated in [6], where four additional axioms are taken into account to construct a covariant representation of the general theory of relativity based on the Hamiltonian in the DeDonder–Weyl formalism.

Therefore, we will try to systematize different approaches and combine them into a single picture, and will also present our own viewpoint on the problem. In our calculations we will everywhere use the metric signature of the form (+,–,–,–). All obtained results will be illustrated using the example of purely vector fields, including the gravitational field within the framework of the covariant theory of gravity (CTG). In CTG, the Lagrangian of the gravitational field is similar in form to the Lagrangian of the electromagnetic field. In this regard, we will refer to [8], where it is indicated that only Maxwell-type theories can be self-consistent, and that there will be no unphysical states in them from the viewpoint of quantum field theory.

## 2. Methods

In curved spacetime, the element of the four-dimensional volume should be written in a covariant form. As a result, instead of (1) we obtain the following standard expression:

$$S = \int_{t_1}^{t_2} L \, dt = \int_{t_1}^{t_2} \left( \int_V \mathcal{L} \sqrt{-g} \, dx^1 dx^2 dx^3 \right) dt = \frac{1}{c} \int_\Omega \mathcal{L} \, d\Omega, \tag{3}$$

where $d\Omega = \sqrt{-g} \, dx^0 dx^1 dx^2 dx^3$ is the element of the covariant four-volume, $dx^1 dx^2 dx^3$ is the product of the differentials of the space coordinates, $dx^0 = c \, dt$, $c$ is the speed of light, and $g$ is the determinant of the metric tensor $g_{\mu\nu}$.

According to the principle of least action, to find the equations of motion of the particles and fields, the action variation in (3) should be equated to zero:

$$\delta S = \int_{t_1}^{t_2} \delta L \, dt = \int_{t_1}^{t_2} \left( \int_V \delta \left( \mathcal{L} \sqrt{-g} \right) dx^1 dx^2 dx^3 \right) dt = 0. \tag{4}$$

Let us now determine, which variables the Lagrangian density can depend on. In the general case, scalar functions, four-vectors and four-tensors, as well as various four-dimensional derivatives of these quantities can serve as such variables. To speak more specifically, we use the Lagrangian density for the four vector fields as an example, according to [7], [9]:

$$\mathcal{L} = -A_\mu j^\mu - D_\mu J^\mu - U_\mu J^\mu - \pi_\mu J^\mu - \frac{1}{4\mu_0} F_{\mu\nu} F^{\mu\nu} + \frac{c^2}{16\pi G} \Phi_{\mu\nu} \Phi^{\mu\nu} - \frac{c^2}{16\pi \eta} u_{\mu\nu} u^{\mu\nu} - \frac{c^2}{16\pi \sigma} f_{\mu\nu} f^{\mu\nu} + ckR - 2ck\Lambda, \tag{5}$$

where $A_\mu = \left(\dfrac{\varphi}{c}, -\mathbf{A}\right)$ is the four-potential of the electromagnetic field, given by the scalar potential $\varphi$ and the vector potential $\mathbf{A}$ of this field,

$j^\mu = \rho_{0q} u^\mu$ is the charge four-current,

$\rho_{0q}$ is the invariant charge density, determined in the reference frame comoving with the particle or element of matter,

$u^\mu$ is the four-velocity of a point particle or element of matter,

$D_\mu = \left(\dfrac{\psi}{c}, -\mathbf{D}\right)$ is the four-potential of the gravitational field, described in terms of the scalar potential $\psi$ and the vector potential $\mathbf{D}$ within the framework of the covariant theory of gravitation,

$J^\mu = \rho_0 u^\mu$ is the mass four-current,

$\rho_0$ is the invariant mass density, determined in the reference frame comoving with the particle or element of matter,

$U_\mu = \left(\dfrac{\vartheta}{c}, -\mathbf{U}\right)$ is the four-potential of the acceleration field, where $\vartheta$ and $\mathbf{U}$ denote the scalar and vector potentials, respectively,

$\pi_\mu = \left(\dfrac{\wp}{c}, -\mathbf{\Pi}\right)$ is the four-potential of the pressure field, consisting of the scalar potential $\wp$ and the vector potential $\mathbf{\Pi}$,

$\mu_0$ is the magnetic constant,

$F_{\mu\nu} = \nabla_\mu A_\nu - \nabla_\nu A_\mu = \partial_\mu A_\nu - \partial_\nu A_\mu$ is the electromagnetic tensor,

$G$ is the gravitational constant,

$\Phi_{\mu\nu} = \nabla_\mu D_\nu - \nabla_\nu D_\mu = \partial_\mu D_\nu - \partial_\nu D_\mu$ is the gravitational tensor,

$\eta$ is the acceleration field coefficient,

$u_{\mu\nu} = \nabla_\mu U_\nu - \nabla_\nu U_\mu = \partial_\mu U_\nu - \partial_\nu U_\mu$ is the acceleration tensor, calculated as the four-curl of the four-potential of the acceleration field,

$\sigma$ is the pressure field coefficient,

$f_{\mu\nu} = \nabla_\mu \pi_\nu - \nabla_\nu \pi_\mu = \partial_\mu \pi_\nu - \partial_\nu \pi_\mu$ is the pressure field tensor,

$k = -\dfrac{c^3}{16\pi G\beta}$, where $\beta$ is a coefficient of the order of unity to be determined,

$R$ is the scalar curvature,

$\Lambda$ is the cosmological constant.

The invariant charge density $\rho_{0q}$, included in the expression of the charge current $j^\mu = \rho_{0q} u^\mu$, and the invariant mass density $\rho_0$, included in the expression of the mass current $J^\mu = \rho_0 u^\mu$, are determined in the same way as in [10].

The description of gravitation using four-potential $D_\mu$ of the gravitational field significantly distinguishes the covariant theory of gravitation (CTG) from the general theory of relativity, where all gravitational effects are taken into account only through the metric tensor, but $D_\mu$ is absent. According to $D_\mu$ gravity can manifest itself in the CTG as an independent physical force, behaving like an electromagnetic force. In the simplest case, taking into account the metric in CTG leads only to that change in the results of spacetime measurements that occurs under the influence of the fields of the system under consideration.

According to (5), the Lagrangian density depends on the following quantities:

$$\mathcal{L} = \mathcal{L}(j^\mu, J^\mu, A_\mu, D_\mu, U_\mu, \pi_\mu, F_{\mu\nu}, \Phi_{\mu\nu}, u_{\mu\nu}, f_{\mu\nu}, R, g^{\mu\nu}). \qquad (6)$$

We can assume that all the quantities in (6) in one way or another depend on the position vector $x^\mu$ specifying an arbitrary point in the system's spacetime. In addition, some of these quantities depend on the four-velocity $u^\mu$ of the matter particle or the point matter element under consideration, for example, the mass four-current $J^\mu$ and the charge four-current $j^\mu$. We included in (6) the four-potentials and tensors of all the fields, as well as the metric tensor $g^{\mu\nu}$, on which some terms in (4) and (5) depend directly. If the Lagrangian density depends on other variables, they can also be added to (6).

Note that in (6) we use the metric tensor with contravariant indices instead of the tensor with covariant indices. This is possible, since $g_{\mu\nu} g^{\mu\nu} = 4$ and the components of $g^{\mu\nu}$ are functions of the components of $g_{\mu\nu}$, and replacing $g^{\mu\nu}$ with $g_{\mu\nu}$ in (6) will not change the subsequent calculations. Additionally, from the point of view of tensor analysis, the results of tensor operations cannot depend on the choice of tensor indices.

We also point to the difference in the form of the derivatives in (6). For example, the four-velocity is defined as the derivative of position vector of the moving particle with respect to the interval: $u^{\mu} = c \dfrac{dx^{\mu}}{ds}$, where $ds$ denotes the interval. Moreover, the vector field tensor is usually found as the four-curl of the four-potential, that is, using the antisymmetric combination of two covariant derivatives. For the field tensors in (5) and (6) we can write the following:

$$F_{\mu\nu} = \nabla_{\mu} \times A_{\nu} = \nabla_{\mu} A_{\nu} - \nabla_{\nu} A_{\mu}, \qquad \Phi_{\mu\nu} = \nabla_{\mu} \times D_{\nu} = \nabla_{\mu} D_{\nu} - \nabla_{\nu} D_{\mu},$$

$$u_{\mu\nu} = \nabla_{\mu} \times U_{\nu} = \nabla_{\mu} U_{\nu} - \nabla_{\nu} U_{\mu}, \qquad f_{\mu\nu} = \nabla_{\mu} \times \pi_{\nu} = \nabla_{\mu} \pi_{\nu} - \nabla_{\nu} \pi_{\mu}, \qquad (7)$$

Suppose now that all the quantities in (5), except for the metric tensor, are completely expressed in terms of $x^{\mu}$ and $u^{\mu}$; thus, instead of (6), $\mathcal{L} = \mathcal{L}(x^{\mu}, u^{\mu}, g^{\mu\nu})$ will exist. In this case, it is convenient to consider the metric tensor as an independent variable. Then, the action variation with respect to the metric tensor gives an equation that allows us to determine the metric tensor in terms of $x^{\mu}$ and $u^{\mu}$.

To derive the equation of the system's motion in case of the given values of the four-currents, four-potentials, four-tensors and fields as functions of $x^{\mu}$ and $u^{\mu}$, we express action variation (4) as follows:

$$\delta S = \int_{t_1}^{t_2} \int_V \delta \mathcal{L} \sqrt{-g}\, dx^1 dx^2 dx^3 dt + \int_{t_1}^{t_2} \int_V \mathcal{L} \delta \sqrt{-g}\, dx^1 dx^2 dx^3 dt = 0. \qquad (8)$$

The determinant $g$ of the metric tensor depends directly on the metric tensor components, so following, for example, [10], we can write the following:

$$\delta \sqrt{-g} = -\frac{1}{2} \sqrt{-g}\, g_{\mu\nu} \delta g^{\mu\nu}. \qquad (9)$$

For the variation of the action in (8) taking into account (9) the following is obtained:

$$\delta S = \int_{t_1}^{t_2} \int_V \delta \mathcal{L} \sqrt{-g}\, dx^1 dx^2 dx^3 dt - \frac{1}{2} \int_{t_1}^{t_2} \int_V \mathcal{L} g_{\mu\nu} \delta g^{\mu\nu} \sqrt{-g}\, dx^1 dx^2 dx^3 dt = 0. \qquad (10)$$

In (10), we should also take into account the variation of Lagrangian density $\mathcal{L} = \mathcal{L}(x^\mu, u^\mu, g^{\mu\nu})$, expressed in terms of variations of the variable quantities:

$$\delta\mathcal{L} = \frac{\partial \mathcal{L}}{\partial x^\mu} \delta x^\mu + \frac{\partial \mathcal{L}}{\partial u^\mu} \delta u^\mu + \frac{\partial \mathcal{L}}{\partial g^{\mu\nu}} \delta g^{\mu\nu}. \tag{11}$$

In the general case, the derivatives with respect to the four-vectors and the four-tensors during the Lagrangian density variation should be considered as functional derivatives.

## 3. Results
### 3.1. Equation of motion and generalized four-force

In this section we consider what the variation $\delta x^\mu$ in (11) gives. This variation is related to the variation $\delta u^\mu$, and both variations are present only in the first integral in (10). Therefore, we can write the following:

$$\delta S_1 = \int_{t_1}^{t_2} \int_V \left( \frac{\partial \mathcal{L}}{\partial x^\mu} \delta x^\mu + \frac{\partial \mathcal{L}}{\partial u^\mu} \delta u^\mu \right) \sqrt{-g}\, dx^1 dx^2 dx^3 dt = 0. \tag{12}$$

According to [11], equality to zero variation $\delta S_1$ in (12) leads to the following equation:

$$\frac{d}{dt}\left( \frac{\partial \mathcal{L}}{\partial u^\mu} \right) = c\, \frac{\partial (\mathcal{L}/u^0)}{\partial x^\mu}. \tag{13}$$

Expression (13) differs from Euler–Lagrange equation (2) by the fact that on the left-hand side the derivative is taken with respect to time and not with respect to $x^\nu$. Let us determine the volumetric density of the generalized four-momentum

$$\mathcal{P}_\mu = -\frac{\partial \mathcal{L}}{\partial u^\mu}. \tag{14}$$

and express (13) in terms of it:

$$\frac{d\mathcal{P}_\mu}{dt} = \mathcal{F}_\mu = -c\frac{\partial(\mathcal{L}/u^0)}{\partial x^\mu}. \tag{15}$$

Equation (15) represents the equation of the matter's motion, expressed in terms of the derivative of the generalized four-momentum density with respect to time. In addition, we introduced the quantity $\mathcal{F}_\mu$ as the volumetric density of the generalized four-force. In fact, (15) is a generalization of Newton's second law, with the replacement of three-dimensional quantities by four-dimensional quantities.

Equations of motion (13) and (15) were first derived in [11], and the volume density $\mathcal{P}_\mu$ of generalized four-momentum (14) was also determined there. In this case, the condition of constancy of the component $u^0$ of each particle at each time interval during the variation process was used, which was justified by using the concept of typical particles. By definition, typical particles define the basic properties of matter in a physical system; therefore, they are convenient for describing continuously distributed matter. After real particles are replaced with typical particles, physical quantities and equations must be associated specifically with the typical particles.

The properties of typical particles are revealed after averaging the properties of real particles over time and over a small volume surrounding a selected observation point. As a result, the component $u^0$ of a typical particle at the observation point actually becomes constant, at least for a physical system that is in equilibrium or in a state of continuous stationary motion. On the other hand, we can assume that if the variation time is short and comparable to $dt$, the component $u^0$ of each particle does not have time to change significantly. In this case, $\mathcal{P}_\mu$ in (14) will be the instantaneous density of the generalized four-momentum.

To understand the meaning of (13), we substitute the Lagrangian $\mathcal{L}$ from (5) into it, which we divide into two parts such that $\mathcal{L} = \mathcal{L}_p + \mathcal{L}_f$. Since the four-currents are determined by the expressions $J^\mu = \rho_0 u^\mu$, and $j^\mu = \rho_{0q} u^\mu$, the four-velocity $u^\mu$ in $\mathcal{L}_p$ becomes the common multiplier:

$$\mathcal{L}_p = -\left(\rho_{0q} A_\mu + \rho_0 D_\mu + \rho_0 U_\mu + \rho_0 \pi_\mu\right)u^\mu. \tag{16}$$

$$\mathcal{L}_f = -\frac{1}{4\mu_0}F_{\mu\nu}F^{\mu\nu} + \frac{c^2}{16\pi G}\Phi_{\mu\nu}\Phi^{\mu\nu} - \frac{c^2}{16\pi\eta}u_{\mu\nu}u^{\mu\nu} - \frac{c^2}{16\pi\sigma}f_{\mu\nu}f^{\mu\nu} + ckR - 2ck\Lambda.$$

(17)

Let us consider the simplest case, when the fields' four-potentials do not depend on the four-velocity $u^\mu$ of matter elements. Since the field tensors are obtained from the four-potentials with the help of the four-curl, in this case the field tensors and $\mathcal{L}_f$ in general will not depend on $u^\mu$. Let us substitute $\mathcal{L} = \mathcal{L}_p + \mathcal{L}_f$ into (14-15):

$$\mathcal{P}_\mu = \rho_{0q} A_\mu + \rho_0 D_\mu + \rho_0 U_\mu + \rho_0 \pi_\mu. \tag{18}$$

$$\frac{d\mathcal{P}_\mu}{dt} = \frac{d}{dt}\left(\rho_{0q} A_\mu + \rho_0 D_\mu + \rho_0 U_\mu + \rho_0 \pi_\mu\right) = \mathcal{F}_\mu = -c\frac{\partial(\mathcal{L}/u^0)}{\partial x^\mu}. \tag{19}$$

According to (18), all the fields present in the system participate in determining the volumetric density of the generalized momentum $\mathcal{P}_\mu$. In this case $\mathcal{P}_\mu$ would depend only on the fields' four-potentials.

Suppose that the right-hand sides of (15) and (19) are equal to zero in any part of the system under consideration. This is possible, for example, if $\frac{\partial(\mathcal{L}/u^0)}{\partial x^\mu} = 0$ and all the variables in the system depend neither on the time nor on the coordinates, at least on average. In this case, both the density of the generalized four-momentum $\mathcal{P}_\mu$ and the integral of $\mathcal{P}_\mu$ over the system's volume, which gives the total generalized four-momentum, are conserved in the system.

As a rule, this situation is characteristic of closed systems, moving by inertia. The conservation of the generalized four-momentum density is also possible in uniform external fields with special configurations. Thus, in crossed electric and magnetic fields, which are perpendicular to each other, a charged particle can move at a certain constant velocity, depending on the value and direction of the fields, along the axis, which is perpendicular to both fields.

If we define $\mathcal{P}_\mu = (\mathcal{P}_0, -\boldsymbol{\mathcal{P}})$, $\mathcal{F}_\mu = (\mathcal{F}_0, -\boldsymbol{\mathcal{F}})$, where $\boldsymbol{\mathcal{P}}$ is the density of the three-dimensional generalized momentum, and $\boldsymbol{\mathcal{F}}$ is the density of the three-dimensional generalized force, then the space component of the four-dimensional equation (15) is written as follows:

$$\frac{d\mathcal{P}}{dt} = \mathcal{F} = c\nabla(\mathcal{L}/u^0), \tag{20}$$

where in view of (18), $\mathcal{P}$ is expressed in terms of the fields' vector potentials:

$$\mathcal{P} = \rho_{0q}\mathbf{A} + \rho_0\mathbf{D} + \rho_0\mathbf{U} + \rho_0\mathbf{\Pi}. \tag{21}$$

If the quantity $\mathcal{L}/u^0$ does not depend explicitly on the coordinates, the density of the generalized force $\mathcal{F}$ vanishes and the density of the generalized momentum $\mathcal{P}$ in (20-21) is conserved.

Let us now express the time component of the four-dimensional equation (15):

$$\frac{d\mathcal{P}_0}{dt} = \mathcal{F}_0 = -\frac{\partial(\mathcal{L}/u^0)}{\partial t}, \tag{22}$$

where in view of (18), $\mathcal{P}_0$ is expressed in terms of the fields' scalar potentials:

$$\mathcal{P}_0 = \frac{1}{c}\left(\rho_{0q}\varphi + \rho_0\psi + \rho_0\vartheta + \rho_0\wp\right). \tag{23}$$

According to (22), if the partial derivative of the quantity $\mathcal{L}/u^0$ with respect to time is equal to zero, then the time component of the generalized four-force density is also equal to zero, and the time component (23) of the generalized four-momentum density is conserved over time.

We note that the quantity $\mathcal{P}_0$ in (23) is the energy density of the system's particles in the scalar potentials of the vector fields, divided by the speed of light. For vector fields, it is known that if we add the energy of the fields themselves to the energy of particles of the system in scalar potentials of the fields, we obtain the relativistic energy of the entire system [7]. In this case, if the following relation holds:

$$\frac{\partial L}{\partial t} = \frac{\partial}{\partial t}\int_V \mathcal{L}\sqrt{-g}\,dx^1 dx^2 dx^3 = 0, \tag{24}$$

then the relativistic energy of the system is conserved, as a consequence of the symmetry associated with the homogeneity of time. To this end, in (24), it is necessary that the conditions $\frac{\partial \mathcal{L}}{\partial t}=0$ and $\frac{\partial \sqrt{-g}}{\partial t}=0$ are simultaneously satisfied; that is, neither the Lagrangian density nor the metric should depend on the coordinate time. In an equilibrium closed system, we should expect that the quantity $\mathcal{L}/u^0$ does not depend directly on time, and $\mathcal{P}_0$ is conserved in (22). Moreover, we can expect that the energy of the fields themselves is also conserved as part of the relativistic energy of the system.

Within the framework of the special theory of relativity, the relation $\frac{\partial \sqrt{-g}}{\partial t}=0$ is valid. Then, if one type of energy is conserved – either the entire relativistic energy, the energy of particles in the fields' scalar potentials, or the energy from the fields' tensor invariants in the Lagrangian density $\mathcal{L}_f$ (17) – then the other two energies are simultaneously conserved.

Notably, in mechanics, the relation $\frac{d\mathbf{p}}{dt}=\mathbf{f}$ is used, where $\mathbf{p}$ and $\mathbf{f}$ denote the density of the ordinary three-dimensional momentum and the density of the ordinary three-dimensional force, respectively. Let us find the relationship between the force density $\mathbf{f}$ and the generalized force density $\mathcal{F}$ for the case of rectilinear motion of a point particle considered to be a solid body. From (20-21) it follows that:

$$\frac{d}{dt}\left(\rho_{0q}\mathbf{A} + \rho_0\mathbf{D} + \rho_0\mathbf{U} + \rho_0\mathbf{\Pi}\right) = \mathcal{F}. \tag{25}$$

In the case of a free solid point particle, the four-potential of the acceleration field coincides with the four-velocity of the particle, $U_\mu = u_\mu$. Within the framework of the special theory of relativity $u_\mu = (\gamma c, -\gamma \mathbf{v})$, $\gamma$ is the Lorentz factor of the particle, and $\mathbf{v}$ is the velocity of the particle. Consequently, $\mathbf{U} = \gamma \mathbf{v}$, and (25) can be transformed by transposing some terms to the right-hand side:

$$\frac{d(\rho_0\mathbf{U})}{dt} = \frac{d(\gamma\rho_0\mathbf{v})}{dt} = \frac{d\mathbf{p}}{dt} = \mathbf{f} = \mathcal{F} - \frac{d}{dt}\left(\rho_{0q}\mathbf{A} + \rho_0\mathbf{D} + \rho_0\mathbf{\Pi}\right). \tag{26}$$

The difference between the force density $\mathbf{f}$ and the generalized force density $\mathcal{F}$ can be seen from the right-hand side of (26). The total time derivative should be considered the material derivative, $\dfrac{d}{dt} = \dfrac{\partial}{\partial t} + \mathbf{v} \cdot \nabla$; therefore, the difference between the forces $\mathbf{f}$ and $\mathcal{F}$ is associated with the time derivatives and with the gradients of the vector potentials of the electromagnetic and gravitational fields, as well as the pressure field in the particle matter.

### 3.2. Equation for metric

We divide the Lagrangian density (5) and (16-17) used by us into two parts such that the first part $\mathcal{L}_p$ contains only the terms with the four-currents, and the second part $\mathcal{L}_f$ explicitly contains the metric tensor among its terms:

$$\mathcal{L}_p = -A_\mu j^\mu - D_\mu J^\mu - U_\mu J^\mu - \pi_\mu J^\mu.$$

$$\begin{aligned}\mathcal{L}_f = &-\frac{1}{4\mu_0} F_{\mu\nu} F_{\eta\lambda} g^{\mu\eta} g^{\lambda\nu} + \frac{c^2}{16\pi G} \Phi_{\mu\nu} \Phi_{\eta\lambda} g^{\mu\eta} g^{\lambda\nu} - \\ &- \frac{c^2}{16\pi \eta} u_{\mu\nu} u_{\eta\lambda} g^{\mu\eta} g^{\lambda\nu} - \frac{c^2}{16\pi \sigma} f_{\mu\nu} f_{\eta\lambda} g^{\mu\eta} g^{\lambda\nu} + c k R_{\mu\nu} g^{\mu\nu} - 2 c k \Lambda.\end{aligned} \quad (27)$$

Using in (10) instead of $\delta \mathcal{L}$ the last term $\dfrac{\partial \mathcal{L}}{\partial g^{\mu\nu}} \delta g^{\mu\nu}$ in (11), we write the variation of the action with respect to the metric tensor:

$$\delta S_2 = \int_{t_1}^{t_2}\!\!\int_V \frac{\partial(\mathcal{L}\sqrt{-g})}{\partial g^{\mu\nu}} \delta g^{\mu\nu}\, dx^1 dx^2 dx^3 dt = \int_{t_1}^{t_2}\!\!\int_V \left( \frac{\partial \mathcal{L}}{\partial g^{\mu\nu}} - \frac{1}{2} \mathcal{L}\, g_{\mu\nu} \right) \delta g^{\mu\nu} \sqrt{-g}\, dx^1 dx^2 dx^3 dt = 0.$$

(28)

In (28) $\mathcal{L} = \mathcal{L}_p + \mathcal{L}_f$ as well as $\dfrac{\partial(\mathcal{L}\sqrt{-g})}{\partial g^{\mu\nu}} = \dfrac{\partial(\mathcal{L}_p\sqrt{-g})}{\partial g^{\mu\nu}} + \dfrac{\partial(\mathcal{L}_f\sqrt{-g})}{\partial g^{\mu\nu}}$.

There are two different possible approaches for analyzing the functional derivative $\dfrac{\partial(\mathcal{L}_p\sqrt{-g})}{\partial g^{\mu\nu}}$. In the first approach it is assumed that the independent variables in the Lagrangian density are the fields' four-potentials, the four-currents and the metric tensor. Then, in the case of variation with respect to the metric tensor, the four-potentials and the four-currents are not varied and act as constants, just like when the partial derivative is taken with respect to one variable, the other variables are considered to be constant. This means that $\frac{\partial \mathcal{L}_p}{\partial g^{\mu\nu}} = 0$, and then $\frac{\partial \left( \mathcal{L}_p \sqrt{-g} \right)}{\partial g^{\mu\nu}} = \mathcal{L}_p \frac{\partial \sqrt{-g}}{\partial g^{\mu\nu}} \neq 0$. The application of the principle of least action allows us to derive equations for determining the four-potentials and the metric tensor (the field equations and equations for the metric) in the physical system under consideration as functions of the position vector $x^\mu$. The four-currents are also found from the equations of motion as certain functions of $x^\mu$. Thus, the fields' four-potentials, four-currents and metric tensor are in the same position with respect to the dependence on $x^\mu$. This approach was used in [7], [9-10].

The other approach is based on the fact that the independent variables in the Lagrangian density are the four-potentials, the position vector $x^\mu$ and the metric tensor. In this case it turns out that

$$\frac{\partial \left( \mathcal{L}_p \sqrt{-g} \right)}{\partial g^{\mu\nu}} = -A_\mu \frac{\partial \left( j^\mu \sqrt{-g} \right)}{\partial g^{\mu\nu}} - D_\mu \frac{\partial \left( J^\mu \sqrt{-g} \right)}{\partial g^{\mu\nu}} - U_\mu \frac{\partial \left( J^\mu \sqrt{-g} \right)}{\partial g^{\mu\nu}} - \pi_\mu \frac{\partial \left( J^\mu \sqrt{-g} \right)}{\partial g^{\mu\nu}} = 0,$$

(29)

since both the four-potentials according to [10] and the products $j^\mu \sqrt{-g}$ and $J^\mu \sqrt{-g}$ are assumed to be not directly dependent on the metric tensor $g^{\mu\nu}$. In this case, $\delta S_2$ in (28) will depend only on $\mathcal{L}_f$, and in the equation for the metric the terms, containing the products of the four-potentials by the four-currents, will vanish. For this reason, in [12], the continuity equations for the four-currents are indicated to be in the form

$$\nabla_\mu j^\mu = \frac{1}{\sqrt{-g}} \partial_\mu \left( \sqrt{-g}\, j^\mu \right) = 0, \qquad \nabla_\mu J^\mu = \frac{1}{\sqrt{-g}} \partial_\mu \left( \sqrt{-g}\, J^\mu \right) = 0, \qquad (30)$$

completely define the four-currents $j^\mu$ and $J^\mu$ at each point of the flow lines in terms of the initial values of these four-currents. As expected, this should lead to products $j^\mu \sqrt{-g}$ and $J^\mu \sqrt{-g}$ not changing when the metric is varied, after which their partial derivatives with respect to the metric tensor vanish.

In this regard, it should be argued that the above continuity equations (30) show that only the products $j^\mu \sqrt{-g}$ and $J^\mu \sqrt{-g}$ do not depend on the relevant coordinates. The dependence of these products on the metric is uncertain.

Furthermore, we follow the first approach with action variation with respect to the metric tensor to derive the equation for the metric. Therefore, according to (28) the equation for the metric has the following form:

$$\frac{\partial \mathcal{L}}{\partial g^{\mu\nu}} = \frac{1}{2} \mathcal{L} g_{\mu\nu}. \tag{31}$$

In (27), the metric tensor is present in the four terms with the field tensors, as well as in the term with the Ricci tensor $R_{\mu\nu}$, the contraction of which with the metric tensor gives the scalar curvature: $R_{\mu\nu} g^{\mu\nu} = R$. We should take into account that the metric tensor appears twice as often as the field tensors. This is why the variation with respect to the metric tensor, for example, for the term with the electromagnetic field, is as follows:

$$\delta \mathcal{L}_{em} = -\frac{1}{4\mu_0} F_{\mu\nu} F_{\eta\lambda} \delta\left(g^{\mu\eta} g^{\lambda\nu}\right) = -\frac{1}{4\mu_0} F_{\mu\nu} F_{\eta\lambda} g^{\mu\eta} \delta g^{\lambda\nu} - \frac{1}{4\mu_0} F_{\mu\nu} F_{\eta\lambda} g^{\lambda\nu} \delta g^{\mu\eta} =$$
$$= -\frac{1}{4\mu_0} F_{\mu\nu} F^\mu{}_\lambda \delta g^{\lambda\nu} - \frac{1}{4\mu_0} F_{\nu\mu} F^\nu{}_\eta \delta g^{\mu\eta} = -\frac{1}{2\mu_0} F_{\mu\nu} F^\mu{}_\lambda \delta g^{\lambda\nu} = \frac{1}{2\mu_0} F_{\nu\alpha} F^\alpha{}_\mu \delta g^{\mu\nu}. \tag{32}$$

The multiplier before $\delta g^{\mu\nu}$ on the right-hand side of (32) is equal to the functional derivative $\dfrac{\partial \mathcal{L}_{em}}{\partial g^{\mu\nu}}$ with respect to the electromagnetic field. Similar variations are present in the terms with the tensors of other fields. With this in mind, we have in (31) the following:

$$\frac{\partial \mathcal{L}}{\partial g^{\mu\nu}} = \frac{1}{2\mu_0} F_{\nu\alpha} F^\alpha{}_\mu - \frac{c^2}{8\pi G} \Phi_{\nu\alpha} \Phi^\alpha{}_\mu + \frac{c^2}{8\pi \eta} u_{\nu\alpha} u^\alpha{}_\mu + \frac{c^2}{8\pi \sigma} f_{\nu\alpha} f^\alpha{}_\mu + ck R_{\mu\nu}. \tag{33}$$

In derivation of (33), we used the fact that the functional derivative of the scalar curvature $R = R_{\mu\nu} g^{\mu\nu}$ with respect to the metric tensor gives $R_{\mu\nu}$ [13]. This is due to the properties of the Ricci tensor $R_{\mu\nu}$ itself.

Let us now substitute $\dfrac{\partial \mathcal{L}}{\partial g^{\mu\nu}}$ (33) into (31) and take into account the Lagrangian density $\mathcal{L}$ in the form of (5):

$$\frac{1}{2\mu_0} F_{\nu\alpha} F^{\alpha}{}_{\mu} - \frac{c^2}{8\pi G} \Phi_{\nu\alpha} \Phi^{\alpha}{}_{\mu} + \frac{c^2}{8\pi \eta} u_{\nu\alpha} u^{\alpha}{}_{\mu} + \frac{c^2}{8\pi \sigma} f_{\nu\alpha} f^{\alpha}{}_{\mu} + c k R_{\mu\nu} =$$

$$= \frac{1}{2} g_{\mu\nu} \left( \begin{array}{c} -A_\alpha\, j^\alpha - D_\alpha\, J^\alpha - U_\alpha\, J^\alpha - \pi_\alpha\, J^\alpha - \dfrac{1}{4\mu_0} F_{\alpha\beta} F^{\alpha\beta} + \dfrac{c^2}{16\pi G} \Phi_{\alpha\beta} \Phi^{\alpha\beta} - \\ -\dfrac{c^2}{16\pi \eta} u_{\alpha\beta} u^{\alpha\beta} - \dfrac{c^2}{16\pi \sigma} f_{\alpha\beta} f^{\alpha\beta} + c k R - 2 c k \Lambda \end{array} \right).$$

(34)

The form of equation (34) can be simplified by using the definitions of the stress-energy tensors of the fields. The standard expression for the stress-energy tensor of the electromagnetic field, which is symmetric with respect to its indices, is as follows:

$$W_{\mu\nu} = \frac{1}{\mu_0} \left( F_{\mu\alpha} F^{\alpha}{}_{\nu} + \frac{1}{4} g_{\mu\nu} F_{\alpha\beta} F^{\alpha\beta} \right). \tag{35}$$

The stress-energy tensors of other vector fields are expressed in the same form. In particular, for the gravitational field within the framework of the covariant theory of gravitation [7], for the acceleration field and for the pressure field, we can write:

$$U_{\mu\nu} = -\frac{c^2}{4\pi G} \left( \Phi_{\mu\alpha} \Phi^{\alpha}{}_{\nu} + \frac{1}{4} g_{\mu\nu} \Phi_{\alpha\beta} \Phi^{\alpha\beta} \right), \qquad B_{\mu\nu} = \frac{c^2}{4\pi \eta} \left( u_{\mu\alpha} u^{\alpha}{}_{\nu} + \frac{1}{4} g_{\mu\nu} u_{\alpha\beta} u^{\alpha\beta} \right),$$

$$P_{\mu\nu} = \frac{c^2}{4\pi \sigma} \left( f_{\mu\alpha} f^{\alpha}{}_{\nu} + \frac{1}{4} g_{\mu\nu} f_{\alpha\beta} f^{\alpha\beta} \right). \tag{36}$$

Considering (35-36), in (34) we obtain the following:

$$2ckR_{\mu\nu} - ckRg_{\mu\nu} = -g_{\mu\nu}\left(A_\alpha j^\alpha + D_\alpha J^\alpha + U_\alpha J^\alpha + \pi_\alpha J^\alpha + 2ck\Lambda\right) - T_{\mu\nu}, \quad (37)$$

where $T_{\mu\nu} = W_{\mu\nu} + U_{\mu\nu} + B_{\mu\nu} + P_{\mu\nu}$ is the total stress-energy tensor of the system.

Let us multiply both sides of equation (37) by $g^{\mu\nu}$ and take into account that $g_{\mu\nu}g^{\mu\nu} = 4$, $R_{\mu\nu}g^{\mu\nu} = R$, and $T_{\mu\nu}g^{\mu\nu} = 0$, since the relations $W_{\mu\nu}g^{\mu\nu} = 0$, $U_{\mu\nu}g^{\mu\nu} = 0$, $B_{\mu\nu}g^{\mu\nu} = 0$, and $P_{\mu\nu}g^{\mu\nu} = 0$ are valid:

$$ckR = 2\left(A_\alpha j^\alpha + D_\alpha J^\alpha + U_\alpha J^\alpha + \pi_\alpha J^\alpha + 2ck\Lambda\right). \quad (38)$$

According to [7], to gauge the relativistic energy of the system, the cosmological constant is given by:

$$-ck\Lambda = A_\mu j^\mu + D_\mu J^\mu + U_\mu J^\mu + \pi_\mu J^\mu. \quad (39)$$

Substitution of (39) into (38) and (37) with $k = -\dfrac{c^3}{16\pi G\beta}$, where $\beta$ is a constant of the order of unity, gives the following:

$$ckR = -2\left(A_\alpha j^\alpha + D_\alpha J^\alpha + U_\alpha J^\alpha + \pi_\alpha J^\alpha\right) = 2ck\Lambda.$$

$$R_{\mu\nu} - \frac{1}{4}Rg_{\mu\nu} = \frac{8\pi G\beta}{c^4}\left(W_{\mu\nu} + U_{\mu\nu} + B_{\mu\nu} + P_{\mu\nu}\right) = \frac{8\pi G\beta}{c^4}T_{\mu\nu}. \quad (40)$$

Equation (40) represents the equation for the metric and has the same form as in [7]. This equation allows us to find the metric tensor components with a given stress-energy tensor $T_{\mu\nu} = W_{\mu\nu} + U_{\mu\nu} + B_{\mu\nu} + P_{\mu\nu}$.

The tensor $T_{\mu\nu}$ consists of the sum of the stress-energy tensors of all the vector fields present in the system and is taken into account in the Lagrangian density. We can single out in Lagrangian density (5) the part that contains the fields' tensor invariants:

$$\mathcal{L}_t = -\frac{1}{4\mu_0} F_{\alpha\beta} F^{\alpha\beta} + \frac{c^2}{16\pi G} \Phi_{\alpha\beta} \Phi^{\alpha\beta} - \frac{c^2}{16\pi \eta} u_{\alpha\beta} u^{\alpha\beta} - \frac{c^2}{16\pi \sigma} f_{\alpha\beta} f^{\alpha\beta}. \qquad (41)$$

The tensor $T_{\mu\nu}$ is expressed in terms of $\mathcal{L}_t$ (41) by the following formula:

$$T_{\mu\nu} = 2 \frac{\partial \mathcal{L}_t}{\partial g^{\mu\nu}} - \mathcal{L}_t g_{\mu\nu} = \frac{2}{\sqrt{-g}} \frac{\partial (\mathcal{L}_t \sqrt{-g})}{\partial g^{\mu\nu}}. \qquad (42)$$

The right-hand side of (42) takes into account the relation $d\sqrt{-g} = -\frac{1}{2}\sqrt{-g}\, g_{\mu\nu} dg^{\mu\nu}$.

In the general theory of relativity, a similar formula is used to determine the Hilbert stress-energy tensor as the total stress-energy tensor of the matter and nongravitational fields. In this case, instead of the Lagrangian density $\mathcal{L}_t$, part of the Lagrangian density is substituted into (42), which in the general theory of relativity does not contain the scalar curvature or the cosmological constant and is responsible for the matter and nongravitational fields [14-15].

### 3.3. Field equations

Let us write the dependence on the variables for the Lagrangian density (5) in the following form:

$$\mathcal{L} = \mathcal{L}\left[ j^\mu, J^\mu, A_\mu, D_\mu, U_\mu, \pi_\mu, F_{\mu\nu}(A_\mu), \Phi_{\mu\nu}(D_\mu), u_{\mu\nu}(U_\mu), f_{\mu\nu}(\pi_\mu), g^{\mu\nu}, R(g^{\mu\nu}) \right]. \qquad (43)$$

We can assume that four-currents, four-potentials, field tensors and the metric tensor in (43) exist independently and therefore must vary independently of each other. The variation of Lagrangian density (43) is as follows:

$$\delta \mathcal{L} = \frac{\partial \mathcal{L}}{\partial j^\mu} \delta j^\mu + \frac{\partial \mathcal{L}}{\partial J^\mu} \delta J^\mu + \frac{\partial \mathcal{L}}{\partial A_\mu} \delta A_\mu + \frac{\partial \mathcal{L}}{\partial D_\mu} \delta D_\mu + \frac{\partial \mathcal{L}}{\partial U_\mu} \delta U_\mu + \frac{\partial \mathcal{L}}{\partial \pi_\mu} \delta \pi_\mu +$$
$$+ \frac{\partial \mathcal{L}}{\partial F_{\mu\nu}} \delta F_{\mu\nu} + \frac{\partial \mathcal{L}}{\partial \Phi_{\mu\nu}} \delta \Phi_{\mu\nu} + \frac{\partial \mathcal{L}}{\partial u_{\mu\nu}} \delta u_{\mu\nu} + \frac{\partial \mathcal{L}}{\partial f_{\mu\nu}} \delta f_{\mu\nu} + \frac{\partial \mathcal{L}}{\partial g_{\mu\nu}} \delta g_{\mu\nu}.$$

(44)

We use (44) to derive the equations of the electromagnetic field. In (44) only two terms are related to the four-potential $A_\mu$, namely $\dfrac{\partial \mathcal{L}}{\partial A_\mu} \delta A_\mu$ and $\dfrac{\partial \mathcal{L}}{\partial F_{\mu\nu}} \delta F_{\mu\nu}$. The action variation associated with the variation of the four-potential $A_\mu$ must vanish. We will substitute the abovementioned two terms into the first integral in (10):

$$\delta S_3 = \int_{t_1}^{t_2} \int_V \left( \frac{\partial \mathcal{L}}{\partial A_\mu} \delta A_\mu + \frac{\partial \mathcal{L}}{\partial F_{\mu\nu}} \delta F_{\mu\nu} \right) \sqrt{-g}\, dx^1 dx^2 dx^3 dt = 0. \tag{45}$$

According to (7) $F_{\mu\nu} = \nabla_\mu \times A_\nu = \nabla_\mu A_\nu - \nabla_\nu A_\mu$; therefore $\delta F_{\mu\nu} = \delta(\nabla_\mu \times A_\nu) = \nabla_\mu \times \delta A_\nu$. We use this for transformation in (45):

$$\begin{aligned}
&\frac{\partial \mathcal{L}}{\partial A_\mu} \delta A_\mu + \frac{\partial \mathcal{L}}{\partial F_{\mu\nu}} \nabla_\mu \times \delta A_\nu = \\
&= \frac{\partial \mathcal{L}}{\partial A_\mu} \delta A_\mu + \nabla_\mu \left( \frac{\partial \mathcal{L}}{\partial F_{\mu\nu}} \delta A_\nu \right) - \nabla_\nu \left( \frac{\partial \mathcal{L}}{\partial F_{\mu\nu}} \delta A_\mu \right) - \nabla_\mu \left( \frac{\partial \mathcal{L}}{\partial F_{\mu\nu}} \right) \delta A_\nu + \nabla_\nu \left( \frac{\partial \mathcal{L}}{\partial F_{\mu\nu}} \right) \delta A_\mu.
\end{aligned} \tag{46}$$

The integral of the difference of the second and third terms on the right-hand side of (46) taken over the four-volume is equal to zero. This term is equal to the antisymmetric difference between two divergences, which becomes the difference between two partial derivatives of the following form:

$$\begin{aligned}
&\int_{t_1}^{t_2} \int_V \left[ \nabla_\mu \left( \frac{\partial \mathcal{L}}{\partial F_{\mu\nu}} \delta A_\nu \right) - \nabla_\nu \left( \frac{\partial \mathcal{L}}{\partial F_{\mu\nu}} \delta A_\mu \right) \right] \sqrt{-g}\, dx^1 dx^2 dx^3 dt = \\
&= \int_{t_1}^{t_2} \int_V \left[ \partial_\mu \left( \frac{\partial \mathcal{L}}{\partial F_{\mu\nu}} \delta A_\nu \sqrt{-g} \right) - \partial_\nu \left( \frac{\partial \mathcal{L}}{\partial F_{\mu\nu}} \delta A_\mu \sqrt{-g} \right) \right] dx^1 dx^2 dx^3 dt.
\end{aligned} \tag{47}$$

We apply the divergence theorem to the first term on the right-hand side of (47):

$$\int_{t_1}^{t_2}\int_V \partial_\mu\left(\frac{\partial \mathcal{L}}{\partial F_{\mu\nu}}\delta A_\nu \sqrt{-g}\right)dx^1 dx^2 dx^3 dt = \frac{1}{c}\left(\int_V \frac{\partial \mathcal{L}}{\partial F_{0\nu}}\delta A_\nu \sqrt{-g}\, dx^1 dx^2 dx^3\right)\Bigg|_{t_1}^{t_2} +$$
$$+\int_{t_1}^{t_2}\oint_\Sigma \frac{\partial \mathcal{L}}{\partial F_{j\nu}}\delta A_\nu\, n_j \sqrt{-g}\, d\Sigma\, dt = 0. \tag{48}$$

The three-dimensional unit vector $n_j$, where the index $j=1,2,3$, represents an outward-directed normal vector to the two-dimensional surface $\Sigma$, surrounding our moving physical system. The equality to zero in (48) follows from the fact that the variation $\delta A_\nu$ of the four-potential of the electromagnetic field at the time points $t_1$ and $t_2$ is equal to zero according to the condition of variation of the action function. In addition, in the case of integration over the surface $\Sigma$, the variation $\delta A_\nu$ on this surface is also considered to equal zero.

For the second term on the right-hand side of (47) we obtain a similar result, therefore it suffices to substitute into (45) only the first, fourth and fifth terms on the right-hand side of (46):

$$\delta S_3 = \int_{t_1}^{t_2}\int_V\left[\frac{\partial \mathcal{L}}{\partial A_\mu}\delta A_\mu - \nabla_\mu\left(\frac{\partial \mathcal{L}}{\partial F_{\mu\nu}}\right)\delta A_\nu + \nabla_\nu\left(\frac{\partial \mathcal{L}}{\partial F_{\mu\nu}}\right)\delta A_\mu\right]\sqrt{-g}\, dx^1 dx^2 dx^3 dt = 0.$$

(49)

In (49), we transform the second term, first changing the places of the indices $\mu$ and $\nu$, and then using the antisymmetry of the electromagnetic field tensor in the form $F_{\mu\nu}=-F_{\nu\mu}$:

$$-\nabla_\mu\left(\frac{\partial \mathcal{L}}{\partial F_{\mu\nu}}\right)\delta A_\nu = -\nabla_\nu\left(\frac{\partial \mathcal{L}}{\partial F_{\nu\mu}}\right)\delta A_\mu = \nabla_\nu\left(\frac{\partial \mathcal{L}}{\partial F_{\mu\nu}}\right)\delta A_\mu.$$

$$\delta S_3 = \int_{t_1}^{t_2}\int_V\left[\frac{\partial \mathcal{L}}{\partial A_\mu} + 2\nabla_\nu\left(\frac{\partial \mathcal{L}}{\partial F_{\mu\nu}}\right)\right]\delta A_\mu \sqrt{-g}\, dx^1 dx^2 dx^3 dt = 0. \tag{50}$$

The Euler–Lagrange equation for the electromagnetic field follows from the above:

$$\frac{\partial \mathcal{L}}{\partial A_\mu} + 2\nabla_\nu \left( \frac{\partial \mathcal{L}}{\partial F_{\mu\nu}} \right) = 0. \qquad (51)$$

Let us substitute into (51) the components of the Lagrangian (27), where the tensor $F_{\mu\nu}$ is contained in $\mathcal{L}_f$ only in one term. The variation in this term with respect to the given tensor will equal:

$$-\frac{1}{4\mu_0}\delta\left(F_{\mu\nu}F_{\eta\lambda}\right)g^{\mu\eta}g^{\lambda\nu} = -\frac{1}{4\mu_0}F_{\mu\nu}g^{\mu\eta}g^{\lambda\nu}\delta F_{\eta\lambda} - \frac{1}{4\mu_0}F_{\eta\lambda}g^{\mu\eta}g^{\lambda\nu}\delta F_{\mu\nu} =$$

$$= -\frac{1}{4\mu_0}F^{\eta\lambda}\delta F_{\eta\lambda} - \frac{1}{4\mu_0}F^{\mu\nu}\delta F_{\mu\nu} = -\frac{1}{2\mu_0}F^{\mu\nu}\delta F_{\mu\nu}. \qquad (52)$$

Taking into account (27) and (52), the functional derivative of the Lagrangian with respect to the tensor $F_{\mu\nu}$ will be as follows: $\frac{\partial \mathcal{L}}{\partial F_{\mu\nu}} = \frac{\partial \mathcal{L}_f}{\partial F_{\mu\nu}} = -\frac{1}{2\mu_0}F^{\mu\nu}$. With the help of $\mathcal{L}_p$ in (27) we find the derivative with respect to the electromagnetic four-potential: $\frac{\partial \mathcal{L}}{\partial A_\mu} = \frac{\partial \mathcal{L}_p}{\partial A_\mu} = -j^\mu$. Substituting all this into (51), we arrive at the equation of the electromagnetic field, to which we can also add the second equation without the field source, resulting from the antisymmetry of the tensor $F_{\mu\nu}$:

$$\nabla_\nu F^{\mu\nu} = -\mu_0 j^\mu, \qquad \nabla_\sigma F_{\mu\nu} + \nabla_\nu F_{\sigma\mu} + \nabla_\mu F_{\nu\sigma} = 0. \qquad (53)$$

Repeating similar steps for other fields, we find equations for the gravitational field, the acceleration field and the pressure field, which are considered vector fields:

$$\nabla_\nu \Phi^{\mu\nu} = \frac{4\pi G}{c^2}J^\mu, \qquad \nabla_\sigma \Phi_{\mu\nu} + \nabla_\nu \Phi_{\sigma\mu} + \nabla_\mu \Phi_{\nu\sigma} = 0,$$

$$\nabla_\nu u^{\mu\nu} = -\frac{4\pi \eta}{c^2}J^\mu, \qquad \nabla_\sigma u_{\mu\nu} + \nabla_\nu u_{\sigma\mu} + \nabla_\mu u_{\nu\sigma} = 0,$$

$$\nabla_\nu f^{\mu\nu} = -\frac{4\pi\sigma}{c^2} J^\mu, \qquad \nabla_\sigma f_{\mu\nu} + \nabla_\nu f_{\sigma\mu} + \nabla_\mu f_{\nu\sigma} = 0. \qquad (54)$$

The second equations in (53-54) for each field hold identically and are the consequence of the fact that the field tensors are defined in terms of the four-curls of the fields' four-potentials [7], [10].

### 3.4. Equation of motion with field tensors

In (19) we found the equation of motion of the physical system, expressed in terms of the time derivative of the fields' four-potentials. However, in addition to the four-potentials, the Lagrangian density also contains field tensors, and it is often more convenient to express the equation of motion precisely in terms of field tensors. This means that instead of the field potentials we can use the field strengths.

Let us take in the Lagrangian density variation (44) the part that is related to the variations of the four-currents, and substitute this part into the first integral in (10):

$$\delta S_4 = \int_{t_1}^{t_2}\int_V \left( \frac{\partial \mathcal{L}}{\partial j^\mu} \delta j^\mu + \frac{\partial \mathcal{L}}{\partial J^\mu} \delta J^\mu \right) \sqrt{-g}\, dx^1 dx^2 dx^3 dt = 0. \qquad (55)$$

Since $j^\mu = \rho_{0q} u^\mu$, $J^\mu = \rho_0 u^\mu$, then the variations of the four-currents are related to the variations of the invariant charge density $\rho_{0q}$, the invariant mass density $\rho_0$ and the four-velocity $u^\mu$:

$$\delta j^\mu = u^\mu \delta\rho_{0q} + \rho_{0q} \delta u^\mu, \qquad \delta J^\mu = u^\mu \delta\rho_0 + \rho_0 \delta u^\mu. \qquad (56)$$

As a rule, the following gauge conditions are imposed on the four-currents: $\nabla_\mu j^\mu = 0$, and $\nabla_\mu J^\mu = 0$, so that the covariant divergences of the four-currents are equal to zero. The gauge conditions represent the so-called continuity equations, relating the charge density (the mass density) and the four-velocity of the moving matter element. Due to this relation, the variations $\delta\rho_{0q}$ and $\delta u^\mu$, $\delta\rho_0$ and $\delta u^\mu$ in (56) become dependent on each other.

On the other hand, we can assume that all the variations $\delta\rho_{0q}$, $\delta\rho_0$ and $\delta u^\mu$ are caused by the same variation $\delta x^\mu$, since the charge density (the mass density) and the four-velocity can be functions of the position vector $x^\mu$, that is, the functions of the coordinates and time. Consequently, in (56) the variations of the four-currents can be expressed in terms of $\delta x^\mu$. When the matter element is moving, the velocity of its motion and the charge density (the mass density) can change, but the charge and mass of this matter are, as a rule, invariants of motion and are conserved. This condition was used in [10], [13] to define the variations of the four-currents in terms of the variations $\delta x^\mu$:

$$\delta j^\mu = \nabla_\sigma \left( j^\sigma \delta x^\mu - j^\mu \delta x^\sigma \right) = \frac{1}{\sqrt{-g}} \partial_\sigma \left[ \sqrt{-g} \left( j^\sigma \delta x^\mu - j^\mu \delta x^\sigma \right) \right],$$

$$\delta J^\mu = \nabla_\sigma \left( J^\sigma \delta x^\mu - J^\mu \delta x^\sigma \right) = \frac{1}{\sqrt{-g}} \partial_\sigma \left[ \sqrt{-g} \left( J^\sigma \delta x^\mu - J^\mu \delta x^\sigma \right) \right]. \quad (57)$$

Using the expression for the Lagrangian density $\mathcal{L} = \mathcal{L}_p + \mathcal{L}_f$ and taking into account (27), we find the functional derivatives with respect to the four-currents:

$$\frac{\partial \mathcal{L}}{\partial j^\mu} = \frac{\partial \mathcal{L}_p}{\partial j^\mu} = -A_\mu, \qquad \frac{\partial \mathcal{L}}{\partial J^\mu} = \frac{\partial \mathcal{L}_p}{\partial J^\mu} = -D_\mu - U_\mu - \pi_\mu. \quad (58)$$

Let us substitute the derivatives (58) and the variations of the four-currents (57) into (55):

$$\delta S_4 = -\int_{t_1}^{t_2} \int_V \left\{ \begin{array}{l} A_\mu \partial_\sigma \left[ \sqrt{-g} \left( j^\sigma \delta x^\mu - j^\mu \delta x^\sigma \right) \right] + \\ + \left( D_\mu + U_\mu + \pi_\mu \right) \partial_\sigma \left[ \sqrt{-g} \left( J^\sigma \delta x^\mu - J^\mu \delta x^\sigma \right) \right] \end{array} \right\} dx^1 dx^2 dx^3 dt = 0. \quad (59)$$

We will transform by parts the first term in (59), related to the charge four-current $j^\mu$:

$$\int_{t_1}^{t_2} \int_V A_\mu \partial_\sigma \left[ \sqrt{-g} \left( j^\sigma \delta x^\mu - j^\mu \delta x^\sigma \right) \right] dx^1 dx^2 dx^3 dt =$$

$$= \int_{t_1}^{t_2} \int_V \partial_\sigma \left\{ A_\mu \left[ \sqrt{-g} \left( j^\sigma \delta x^\mu - j^\mu \delta x^\sigma \right) \right] \right\} dx^1 dx^2 dx^3 dt - \quad (60)$$

$$- \int_{t_1}^{t_2} \int_V \left[ \sqrt{-g} \left( j^\sigma \delta x^\mu - j^\mu \delta x^\sigma \right) \right] \partial_\sigma A_\mu \, dx^1 dx^2 dx^3 dt.$$

The first integral on the right-hand side of (60) is equal to zero, similar to (48), since it is an integral of the divergence of a certain four-vector. The integrand in the last term on the right-hand side of (60) can be transformed by changing the indices:

$$\left[\sqrt{-g}\left(j^\sigma \delta x^\mu - j^\mu \delta x^\sigma\right)\right]\partial_\sigma A_\mu = \left(j^\sigma \delta x^\mu \partial_\sigma A_\mu - j^\sigma \delta x^\mu \partial_\mu A_\sigma\right)\sqrt{-g} =$$
$$= j^\sigma \left(\partial_\sigma A_\mu - \partial_\mu A_\sigma\right)\delta x^\mu \sqrt{-g} = j^\sigma F_{\sigma\mu} \delta x^\mu \sqrt{-g}.$$

(61)

Repeating the same as in (60-61), for the gravitational field, the acceleration field and the pressure field with respect to the mass four-current $J^\mu$, for (59) we find the following:

$$\delta S_4 = \int_{t_1}^{t_2}\int_V \left[j^\sigma F_{\sigma\mu} + J^\sigma\left(\Phi_{\sigma\mu} + u_{\sigma\mu} + f_{\sigma\mu}\right)\right]\delta x^\mu \sqrt{-g}\, dx^1 dx^2 dx^3 dt = 0. \qquad (62)$$

From (62) it follows, similarly to [7]. [9], [16-17], equation of motion of charged matter in four vector fields:

$$j^\sigma F_{\sigma\mu} + J^\sigma\left(\Phi_{\sigma\mu} + u_{\sigma\mu} + f_{\sigma\mu}\right) = 0. \qquad (63)$$

The equation of motion (63) can be derived without variation with respect to the variable $x^\mu$. If we substitute (39) into (37), then the equation for the metric will have the following form:

$$2ckR_{\mu\nu} - ckRg_{\mu\nu} = 2ckG_{\mu\nu} = -ck\Lambda g_{\mu\nu} - T_{\mu\nu}. \qquad (64)$$

On the left-hand side of (64), the Einstein tensor $G_{\mu\nu} = R_{\mu\nu} - \dfrac{1}{2}Rg_{\mu\nu}$ is multiplied by $2ck$. Since the divergence of this tensor is equal to zero: $\nabla^\nu G_{\mu\nu} = 0$, the divergence of the right-hand side will also be equal to zero:

$$-ck\nabla_\mu \Lambda - \nabla^\nu T_{\mu\nu} = 0. \qquad (65)$$

The cosmological constant $\Lambda$ is constant in the sense that, for it, the following holds true $\nabla_\mu \Lambda = 0$. Therefore, in (65) the relation must be satisfied

$$\nabla^\nu T_{\mu\nu} = \nabla^\nu \left( W_{\mu\nu} + U_{\mu\nu} + B_{\mu\nu} + P_{\mu\nu} \right) = 0 . \tag{66}$$

Substituting into (66) the stress-energy tensors of fields (35-36), and taking into account the relations from [7], [9]:

$$\nabla^\nu W_{\mu\nu} = j^\sigma F_{\sigma\mu}, \qquad \nabla^\nu U_{\mu\nu} = J^\sigma \Phi_{\sigma\mu},$$

$$\nabla^\nu B_{\mu\nu} = J^\sigma u_{\sigma\mu}, \qquad \nabla^\nu P_{\mu\nu} = J^\sigma f_{\sigma\mu}, \tag{67}$$

we arrive at the equation of motion (63).

In the derivation of equation (63) we used variations (57), which are valid on the condition that the continuity equations are satisfied: $\nabla_\mu j^\mu = 0$, $\nabla_\mu J^\mu = 0$.

Like in [7], we take into account the divergences of field equations (53-54) and the continuity equations. This leads to the following relations:

$$\nabla_\mu \nabla_\nu F^{\mu\nu} = \nabla_\mu \nabla_\nu \nabla^\mu A^\nu - \nabla_\mu \nabla_\nu \nabla^\nu A^\mu = -R_{\mu\nu} F^{\mu\nu} = -\mu_0 \nabla_\mu j^\mu = 0,$$

$$R_{\mu\nu} \Phi^{\mu\nu} = -\frac{4\pi G}{c^2} \nabla_\mu J^\mu = 0, \quad R_{\mu\nu} u^{\mu\nu} = \frac{4\pi \eta}{c^2} \nabla_\mu J^\mu = 0, \quad R_{\mu\nu} f^{\mu\nu} = \frac{4\pi \sigma}{c^2} \nabla_\mu J^\mu = 0.$$

$$\tag{68}$$

According to (68), in curved spacetime an additional relation appears between the Ricci tensor $R_{\mu\nu}$ and the field tensors, so that contraction of $R_{\mu\nu}$ with the field tensors must be equal to zero.

If we take into account the expression of equation of motion (66) and formula (42), then for the equation of motion we obtain the following:

$$\nabla^\nu T_{\mu\nu} = \nabla^\nu \left( 2 \frac{\partial \mathcal{L}_t}{\partial g^{\mu\nu}} - \mathcal{L}_t g_{\mu\nu} \right) = \nabla^\nu \left[ \frac{2}{\sqrt{-g}} \frac{\partial \left( \mathcal{L}_t \sqrt{-g} \right)}{\partial g^{\mu\nu}} \right] = 0 , \qquad (69)$$

where $\mathcal{L}_t$ represents the part of the Lagrangian density, that contains the tensor invariants of the fields and for vector fields has the form (41). In this case, the time component of equation (69) describes the generalized Poynting theorem for energy and energy flows of fields [18].

### 3.5. Relativistic energy

The standard determination of the formula for the system's relativistic energy is carried out in two stages [19]. First, the Euler–Lagrange equation is derived on the assumption that the Lagrangian depends only on the current time $t$, on the three-dimensional radius-vector $\mathbf{r}_n$, which specifies the location of the element of matter of the system with the current number $n$ at the moment of time $t$, and on the velocity $\mathbf{v}_n$ of motion of this element of matter. Thus, for the Lagrangian the dependence is assumed in the form $L = L(t, \mathbf{r}_n, \mathbf{v}_n)$. Next, the action $S$ is varied and the variation $\delta S$ is equated to zero. In this case, the action does not vary with respect to time:

$$S = \int_{t_1}^{t_2} L \, dt , \qquad \delta S = \int_{t_1}^{t_2} \delta L \, dt = \int_{t_1}^{t_2} \sum_{n=1}^{N} \left( \frac{\partial L}{\partial \mathbf{r}_n} \cdot \delta \mathbf{r}_n + \frac{\partial L}{\partial \mathbf{v}_n} \cdot \delta \mathbf{v}_n \right) dt = 0 . \qquad (70)$$

Since $\mathbf{v}_n = \dfrac{d \mathbf{r}_n}{dt}$, in (70) we have

$$\frac{\partial L}{\partial \mathbf{v}_n} \cdot \delta \mathbf{v}_n = \frac{\partial L}{\partial \mathbf{v}_n} \cdot \delta \frac{d \mathbf{r}_n}{dt} = \frac{\partial L}{\partial \mathbf{v}_n} \cdot \frac{d(\delta \mathbf{r}_n)}{dt} = \frac{d}{dt} \left( \frac{\partial L}{\partial \mathbf{v}_n} \cdot \delta \mathbf{r}_n \right) - \frac{d}{dt} \left( \frac{\partial L}{\partial \mathbf{v}_n} \right) \cdot \delta \mathbf{r}_n ,$$

$$\delta S = \int_{t_1}^{t_2} \sum_{n=1}^{N} \left[ \frac{\partial L}{\partial \mathbf{r}_n} - \frac{d}{dt} \left( \frac{\partial L}{\partial \mathbf{v}_n} \right) \right] \cdot \delta \mathbf{r}_n \, dt + \int_{t_1}^{t_2} \sum_{n=1}^{N} \frac{d}{dt} \left( \frac{\partial L}{\partial \mathbf{v}_n} \cdot \delta \mathbf{r}_n \right) dt = 0 . \qquad (71)$$

In this case, for the second term in (71) there is the equality

$$\int_{t_1}^{t_2} \sum_{n=1}^{N} \frac{d}{dt}\left(\frac{\partial L}{\partial \mathbf{v}_n} \cdot \delta \mathbf{r}_n\right) dt = \sum_{n=1}^{N} \left(\frac{\partial L}{\partial \mathbf{v}_n} \cdot \delta \mathbf{r}_n\right)\Bigg|_{t_1}^{t_2} = 0,$$

since the variations $\delta \mathbf{r}_n$ are equal to zero at the time points $t_1$ and $t_2$. Then the Euler–Lagrange equation for each matter element follows from the condition $\delta S = 0$ in (71):

$$\frac{\partial L}{\partial \mathbf{r}_n} - \frac{d}{dt}\left(\frac{\partial L}{\partial \mathbf{v}_n}\right) = 0. \tag{72}$$

In the second stage, we search for the time derivative of the Lagrangian, expressed in terms of the coordinates and the velocity of each of $N$ matter elements:

$$\frac{dL}{dt} = \frac{\partial L}{\partial t} + \sum_{n=1}^{N}\left(\frac{\partial L}{\partial \mathbf{r}_n} \cdot \frac{d\mathbf{r}_n}{dt} + \frac{\partial L}{\partial \mathbf{v}_n} \cdot \frac{d\mathbf{v}_n}{dt}\right) = \frac{\partial L}{\partial t} + \sum_{n=1}^{N}\left(\frac{\partial L}{\partial \mathbf{r}_n} \cdot \mathbf{v}_n + \frac{\partial L}{\partial \mathbf{v}_n} \cdot \frac{d\mathbf{v}_n}{dt}\right). \tag{73}$$

Using in (73) $\frac{\partial L}{\partial \mathbf{r}_n}$ (72), we find:

$$\frac{dL}{dt} = \frac{\partial L}{\partial t} + \sum_{n=1}^{N}\left[\frac{d}{dt}\left(\frac{\partial L}{\partial \mathbf{v}_n}\right) \cdot \mathbf{v}_n + \frac{\partial L}{\partial \mathbf{v}_n} \cdot \frac{d\mathbf{v}_n}{dt}\right] = \frac{\partial L}{\partial t} + \frac{d}{dt}\sum_{n=1}^{N}\left(\frac{\partial L}{\partial \mathbf{v}_n} \cdot \mathbf{v}_n\right).$$

$$\frac{d}{dt}\left[\sum_{n=1}^{N}\left(\frac{\partial L}{\partial \mathbf{v}_n} \cdot \mathbf{v}_n\right) - L\right] = \frac{dE}{dt} = -\frac{\partial L}{\partial t}. \tag{74}$$

We see in (74) that if, in the physical system, the Lagrangian does not depend explicitly on the time and $\frac{\partial L}{\partial t} = 0$, then the system's energy $E$ is conserved, given by the expression:

$$E = \sum_{n=1}^{N}\left(\mathbf{v}_n \cdot \frac{\partial L}{\partial \mathbf{v}_n}\right) - L. \tag{75}$$

Proceeding further as in [20], in (75) one can move from the summation over particles to the integral over the volume of the system, and from the derivative $\dfrac{\partial L}{\partial \mathbf{v}_n}$ to the corresponding derivative of the $\mathcal{L}_p$ in integral over the volume. This gives the following:

$$E = \int_V \left[ \mathbf{v} \cdot \dfrac{\partial}{\partial \mathbf{v}} \left( \dfrac{\mathcal{L}_p}{u^0} \right) - \dfrac{\mathcal{L}_p}{u^0} \right] u^0 \sqrt{-g}\, dx^1 dx^2 dx^3 - \int_V \mathcal{L}_f \sqrt{-g}\, dx^1 dx^2 dx^3 + \sum_{n=1}^{N} \left( \mathbf{v}_n \cdot \dfrac{\partial L_f}{\partial \mathbf{v}_n} \right). \qquad (76)$$

In (76) the expression is provided for the system's relativistic energy, containing the continuously distributed matter. Since $L = L_p + L_f = \int_V (\mathcal{L}_p + \mathcal{L}_f) \sqrt{-g}\, dx^1 dx^2 dx^3$, then taking into account (16-17) for $\mathcal{L}_p$ and $\mathcal{L}_f$ the energy (76) can be represented as follows:

$$E = \dfrac{1}{c} \int_V \left[ \begin{array}{l} -\dfrac{\partial}{\partial \mathbf{v}} \left( \rho_{0q}\varphi + \rho_0 \psi + \rho_0 \vartheta + \rho_0 \wp \right) + \\ +\mathbf{v} \cdot \dfrac{\partial}{\partial \mathbf{v}} \left( \rho_{0q}\mathbf{A} + \rho_0 \mathbf{D} + \rho_0 \mathbf{U} + \rho_0 \mathbf{\Pi} \right) \end{array} \right] \cdot \mathbf{v}\, u^0 \sqrt{-g}\, dx^1 dx^2 dx^3 +$$

$$+ \int_V \left( \begin{array}{l} \dfrac{u^0}{c} \left( \rho_{0q}\varphi + \rho_0 \psi + \rho_0 \vartheta + \rho_0 \wp \right) + \\ + \dfrac{1}{4\mu_0} F_{\mu\nu} F^{\mu\nu} - \dfrac{c^2}{16\pi G} \Phi_{\mu\nu} \Phi^{\mu\nu} + \dfrac{c^2}{16\pi \eta} u_{\mu\nu} u^{\mu\nu} + \dfrac{c^2}{16\pi \sigma} f_{\mu\nu} f^{\mu\nu} \end{array} \right) \sqrt{-g}\, dx^1 dx^2 dx^3 +$$

$$+ \sum_{n=1}^{N} \left( \mathbf{v}_n \cdot \dfrac{\partial L_f}{\partial \mathbf{v}_n} \right).$$

(77)

In (77), the energy calibration condition is taken into account in the form $R = 2\Lambda$, which follows from the first relation in (40).

There are cases, in which the derivatives with respect to the velocity in (77) can be neglected. Suppose the entire physical system is stationary, does not rotate and consists of a multitude of matter elements in the form of typical particles moving randomly at a low velocity. The field strengths and solenoidal vectors at each point in space are found by means of the vector superposition of the fields, generated by all the typical particles. The fields formed in this way can be many times greater than the proper fields generated by an individual typical particle. If this holds true, then at a first approximation we can assume that the field potentials and $L_f$ do not depend on the velocities of individual particles inside the system. Then $\dfrac{\partial L_f}{\partial \mathbf{v}_n} \approx 0$, and the last term in (77) is small in comparison with the other terms. The derivatives of the field potentials with respect to the velocities also vanish. This was used to estimate the energy of the relativistic uniform system in [7] and in [20].

The situation changes, when there are few particles in the system and they move at relativistic velocities, or when the entire system moves at a velocity close to the speed of light. In this case, the last terms in (76-77) can already make a rather significant contribution to the system's energy. The derivatives of the field potentials with respect to the velocities also become significant.

### 3.6. Legendre transformation and Hamiltonian

A mathematical procedure that allows us to pass on from the Lagrangian $L$ to the Hamiltonian $H$ is called the Legendre transformation. The importance of the Hamiltonian lies in its association with the relativistic energy of the physical system. This can be seen from the definition of $H$, which we will write in view of (75-76):

$$H = \sum_{n=1}^{N}\left(\mathbf{v}_n \cdot \frac{\partial L}{\partial \mathbf{v}_n}\right) - L =$$
$$= \int_V \left[\mathbf{v} \cdot \frac{\partial}{\partial \mathbf{v}}\left(\frac{\mathcal{L}_p}{u^0}\right) - \frac{\mathcal{L}_p}{u^0}\right] u^0 \sqrt{-g}\, dx^1 dx^2 dx^3 - \int_V \mathcal{L}_f \sqrt{-g}\, dx^1 dx^2 dx^3 + \sum_{n=1}^{N}\left(\mathbf{v}_n \cdot \frac{\partial L_f}{\partial \mathbf{v}_n}\right).$$
(78)

If the Lagrangian depends explicitly on the time and $\dfrac{\partial L}{\partial t} \neq 0$, then both the Lagrangian and the Hamiltonian vary with time.

For the sake of simplicity, we consider the simplest case, when the mass density $\rho_0$, the charge density $\rho_{0q}$ and the field potentials do not explicitly depend on the particles' velocities. In this case, the tensor invariants and $L_f$ do not depend on the particles' velocities either, so that $\dfrac{\partial L_f}{\partial \mathbf{v}_n} = 0$. Next we need the relation from [2]:

$$\frac{cdt}{ds}\sqrt{-g}\, dx^1 dx^2 dx^3 = \frac{u^0}{c}\sqrt{-g}\, dx^1 dx^2 dx^3 = dV_0,$$
(79)

where $dV_0$ is the volume element in the reference frame associated with the moving matter element, and $\sqrt{-g}\, c\, dt\, dx^1 dx^2 dx^3 = d\Omega$ is the element of the covariant four-volume of this matter from the standpoint of the coordinate observer.

In (16) $\mathcal{L}_p$ can be expressed in terms of the velocity $\mathbf{v}$ of typical particles of the system:

$$\mathcal{L}_p = \frac{u^0}{c}\left(-\rho_{0q}\varphi + \rho_{0q}\mathbf{A}\cdot\mathbf{v} - \rho_0\psi + \rho_0\mathbf{D}\cdot\mathbf{v} - \rho_0\vartheta + \rho_0\mathbf{U}\cdot\mathbf{v} - \rho_0\wp + \rho_0\mathbf{\Pi}\cdot\mathbf{v}\right). \tag{80}$$

Taking into account (79-80) we have:

$$L_p = \int_V \mathcal{L}_p \sqrt{-g}\, dx^1 dx^2 dx^3 = c\int_V \frac{\mathcal{L}_p}{u^0}\, dV_0,$$

$$\mathbf{p}_n = \frac{\partial L}{\partial \mathbf{v}_n} = \frac{\partial(L_p + L_f)}{\partial \mathbf{v}_n} \approx \frac{\partial L_p}{\partial \mathbf{v}_n} = c\int_{V_n} \frac{\partial}{\partial \mathbf{v}}\left(\frac{\mathcal{L}_p}{u^0}\right) dV_0 = \int_{V_n}\left(\rho_{0q}\mathbf{A} + \rho_0\mathbf{D} + \rho_0\mathbf{U} + \rho_0\mathbf{\Pi}\right) dV_0 =$$

$$= \int_{V_n} \mathcal{P}\, dV_0 = \frac{1}{c}\int_{V_n} \mathcal{P}\, u^0 \sqrt{-g}\, dx^1 dx^2 dx^3.$$

$$\tag{81}$$

We recall that $\mathcal{P}$ according to (21) is a consequence of (14), where the derivative of the Lagrangian density $\mathcal{L}$ with respect to the four-velocity $u^\mu$ was calculated, while in (81) $\mathcal{P}$ is found through the derivative with respect to the three-dimensional velocity $\mathbf{v}$ in the form $\mathcal{P} = c\dfrac{\partial}{\partial \mathbf{v}}\left(\dfrac{\mathcal{L}_p}{u^0}\right)$.

In the case under consideration, the quantity $\mathbf{p}_n$ in (81) represents a three-dimensional generalized momentum of one matter element that has the invariant volume $dV_0$ and the velocity $\mathbf{v}_n$.

Substitution of (81) into (78) gives the Legendre transformation for the continuously distributed matter that allows us to pass on from the Lagrangian to the Hamiltonian:

$$H = \sum_{n=1}^{N} \mathbf{v}_n \cdot \mathbf{p}_n - L \approx \int_V \mathcal{P} \cdot \mathbf{v} \, dV_0 - L. \tag{82}$$

The sum of (82) means that the entire system can be divided into $N$ volume elements, each of which can be attributed to its own generalized momentum $\mathbf{p}_n$ and its own velocity $\mathbf{v}_n$. The sum indicated on the right-hand side of (82) is replaced by the integral over the entire volume, occupied by the systems' particles.

In the classical approach, the Hamiltonian $H$ must depend not on the velocity $\mathbf{v}_n$, but on the generalized momentum of the particles and fields of each volume element, $H = H(t, \mathbf{r}_n, \mathbf{p}_n)$. If in (82) we take the differentials of $L = L(t, \mathbf{r}_n, \mathbf{v}_n)$ and $H = H(t, \mathbf{r}_n, \mathbf{p}_n)$, then the following relations hold:

$$dH(t, \mathbf{r}_n, \mathbf{p}_n) = \frac{\partial H}{\partial t} dt + \sum_{n=1}^{N} \frac{\partial H}{\partial \mathbf{r}_n} \cdot d\mathbf{r}_n + \sum_{n=1}^{N} \frac{\partial H}{\partial \mathbf{p}_n} \cdot d\mathbf{p}_n =$$

$$= \sum_{n=1}^{N} \mathbf{v}_n \cdot d\mathbf{p}_n + \sum_{n=1}^{N} \mathbf{p}_n \cdot d\mathbf{v}_n - \frac{\partial L}{\partial t} dt - \sum_{n=1}^{N} \frac{\partial L}{\partial \mathbf{r}_n} \cdot d\mathbf{r}_n - \sum_{n=1}^{N} \frac{\partial L}{\partial \mathbf{v}_n} \cdot d\mathbf{v}_n.$$

$$\frac{\partial H}{\partial t} = -\frac{\partial L}{\partial t}, \qquad \frac{\partial H}{\partial \mathbf{r}_n} = -\frac{\partial L}{\partial \mathbf{r}_n}, \qquad \frac{\partial H}{\partial \mathbf{p}_n} = \mathbf{v}_n, \qquad \frac{\partial L}{\partial \mathbf{v}_n} = \mathbf{p}_n. \tag{83}$$

By substituting the relations $\frac{\partial L}{\partial \mathbf{v}_n} = \mathbf{p}_n$ and $\frac{\partial H}{\partial \mathbf{r}_n} = -\frac{\partial L}{\partial \mathbf{r}_n}$ from (83) into the Euler–Lagrange equation (72), we obtain the following:

$$\frac{d\mathbf{p}_n}{dt} = -\frac{\partial H}{\partial \mathbf{r}_n}. \tag{84}$$

Relations (83-84) represent the standard Hamiltonian equations. In this case, three-dimensional equation (84) repeats the more general four-dimensional equation (15) and can be obtained from the spatial component (20) of this equation. To arrive at (84), it is sufficient to integrate (20) over the invariant volume of one element of matter, taking into account the definition of $\mathbf{p}_n$ in (81), the relation $\dfrac{\partial H}{\partial \mathbf{r}_n} = -\dfrac{\partial L}{\partial \mathbf{r}_n}$ in (83) and the expression for the Lagrangian

$$L = c \int_V \frac{\mathcal{L}}{u^0} dV_0 .$$

### 3.7. Relativistic momentum of a system

In Section 3.5 it was shown that if the Lagrangian does not depend explicitly on time or $\dfrac{\partial L}{\partial t} = 0$ in the physical system under consideration, then the system's energy $E$ is conserved, as a certain additive function. The system's momentum is also an additive function. To determine the momentum and its conservation law, the property of space homogeneity is used, when in the case of simultaneous transfer of all the particles of a closed system to a certain small constant vector $\delta \mathbf{r}$ the state of the system does not change [19].

Variation $\delta \mathbf{r}$ leads to the Lagrangian variation of the following form:

$$\delta L = \sum_{n=1}^{N} \frac{\partial L}{\partial \mathbf{r}_n} \delta \mathbf{r} = \delta \mathbf{r} \sum_{n=1}^{N} \frac{\partial L}{\partial \mathbf{r}_n} . \qquad (85)$$

Due to the vanishing of the action variation in (4), the variation $\delta L$ of the Lagrangian (85) must vanish as well. Since the variation $\delta \mathbf{r}$ is arbitrary, the following condition must hold:

$$\sum_{n=1}^{N} \frac{\partial L}{\partial \mathbf{r}_n} = 0, \qquad (86)$$

where $\mathbf{r}_n$ is the radius-vector of the point in space, where the matter element with number $n$ is located.

We express $\dfrac{\partial L}{\partial \mathbf{r}_n}$ from (72) and substitute into (86):

$$\frac{d}{dt} \sum_{n=1}^{N} \frac{\partial L}{\partial \mathbf{v}_n} = 0 . \qquad (87)$$

Hence, it follows from (87) that in a closed system, the quantity, called the system's momentum, is conserved:

$$\mathbf{P} = \sum_{n=1}^{N} \frac{\partial L}{\partial \mathbf{v}_n} = \sum_{n=1}^{N} \mathbf{P}_n \ . \tag{88}$$

In (88) $\mathbf{P}_n$ is the three-dimensional momentum of one volume element, the Lagrangian is $L = L_p + L_f = \int_V \left( \mathcal{L}_p + \mathcal{L}_f \right) \sqrt{-g}\, dx^1 dx^2 dx^3$. For the case of continuous distribution of matter, it is necessary in (88) to pass on from the sum to the integral, then the system's momentum becomes equal to

$$\mathbf{P} = \int_V \frac{\partial}{\partial \mathbf{v}} \left( \frac{\mathcal{L}_p}{u^0} \right) u^0 \sqrt{-g}\, dx^1 dx^2 dx^3 + \sum_{n=1}^{N} \frac{\partial}{\partial \mathbf{v}_n} \left( \int_V \mathcal{L}_f \sqrt{-g}\, dx^1 dx^2 dx^3 \right), \tag{89}$$

where integration must be performed over the entire system's volume, $\mathcal{L}_p$ and $\mathcal{L}_f$ are components of the Lagrangian density $\mathcal{L}$ of the system.

If in (89) we use expressions for $\mathcal{L}_p$ (80) and $\mathcal{L}_f$ (17), then it is clear that the main contribution to the momentum of the system is made by the quantity $\rho_{0q} \mathbf{A} + \rho_0 \mathbf{D} + \rho_0 \mathbf{U} + \rho_0 \mathbf{\Pi}$, which contains the fields' vector potentials inside the matter and which must be integrated over the volume:

$$\mathbf{P} = \frac{1}{c} \int_V \left\{ \begin{array}{l} \rho_{0q} \mathbf{A} + \rho_0 \mathbf{D} + \rho_0 \mathbf{U} + \rho_0 \mathbf{\Pi} - \frac{\partial}{\partial \mathbf{v}} \left( \rho_{0q} \varphi + \rho_0 \psi + \rho_0 \vartheta + \rho_0 \wp \right) + \\ + \left[ \frac{\partial}{\partial \mathbf{v}} \left( \rho_{0q} \mathbf{A} + \rho_0 \mathbf{D} + \rho_0 \mathbf{U} + \rho_0 \mathbf{\Pi} \right) \right] \cdot \mathbf{v} \end{array} \right\} u^0 \sqrt{-g}\, dx^1 dx^2 dx^3 +$$

$$+ \sum_{n=1}^{N} \frac{\partial}{\partial \mathbf{v}_n} \left( \int_V \mathcal{L}_f \sqrt{-g}\, dx^1 dx^2 dx^3 \right).$$

$$\tag{90}$$

The derivatives $\dfrac{\partial}{\partial \mathbf{v}}$ in (90) take into account the possible dependence of the field potentials on the velocity of the matter elements. In addition, the sum $\sum_{n=1}^{N}\dfrac{\partial L_f}{\partial \mathbf{v}_n} = \sum_{n=1}^{N}\dfrac{\partial}{\partial \mathbf{v}_n}\left(\int_V \mathcal{L}_f \sqrt{-g}\, dx^1 dx^2 dx^3\right)$ should be taken into account in the momentum in case, when part of the Lagrangian density $\mathcal{L}_f$, containing the tensor invariants, depends on the velocities $\mathbf{v}_n$ of the matter particles. This means that the system's momentum is also contributed by the part of the field that goes beyond the matter's limits and moves together with the matter. The momentum $\mathbf{P}$ in the form (90) was also found in [20].

### 3.8. Four-momentum of a system

According to the standard definition, the four-momentum is defined as a four-vector, the components of which include energy and momentum. We should note that in (90), the momentum is formed mainly with the help of the fields' vector potentials inside the matter. These vector potentials are the components of the fields' four-potentials that are covariant four-vectors. In connection with this, as in [20] we define the four-momentum as a covariant vector of the following form:

$$P_\mu = \left(\frac{E}{c},\ -\mathbf{P}\right). \tag{91}$$

As a rule, the motion of a physical system relative to a certain reference frame $K$ is taken into account through the motion of the physical system's center of momentum. By definition, in the center-of-momentum frame $\mathbf{P}=0$, and the energy becomes minimal and equal to the invariant rest energy $E_0$. In such a reference frame $K'$, associated with the center of momentum, there should be

$$P'_\nu = \left(\frac{E_0}{c},\ 0\right). \tag{92}$$

Let us assume that the four-dimensional coordinates $x'^0, x'^1, x'^2, x'^3$ of the reference frame $K'$ are related to the coordinates $x^0, x^1, x^2, x^3$ of the reference frame $K$ by functional relations

$$x'^0 = x'^0\left(x^0, x^1, x^2, x^3\right), \qquad x'^1 = x'^1\left(x^0, x^1, x^2, x^3\right).$$

$$x'^2 = x'^2\left(x^0, x^1, x^2, x^3\right), \qquad x'^3 = x'^3\left(x^0, x^1, x^2, x^3\right).$$

(93)

Then the four-momentum $P'_\nu$ (92) can be converted into four-momentum $P_\mu$ observed in the reference frame $K$ using the formula:

$$P_\mu = P'_\nu S^\nu{}_\mu = P'_\nu \frac{\partial x'^\nu}{\partial x^\mu}. \tag{94}$$

In (94), the components of the Jacobian $S^\nu{}_\mu = \frac{\partial x'^\nu}{\partial x^\mu}$ are calculated using functions (93).

From (91-94) for the time and spatial components of $P_\mu$ the following expressions follow:

$$P_0 = P'_\nu S^\nu{}_0 = P'_\nu \frac{\partial x'^\nu}{\partial x^0} = P'_0 \frac{\partial x'^0}{\partial x^0} + P'_1 \frac{\partial x'^1}{\partial x^0} + P'_2 \frac{\partial x'^2}{\partial x^0} + P'_3 \frac{\partial x'^3}{\partial x^0}.$$

$$P_j = P'_\nu S^\nu{}_j = P'_\nu \frac{\partial x'^\nu}{\partial x^j} = P'_0 \frac{\partial x'^0}{\partial x^j} + P'_1 \frac{\partial x'^1}{\partial x^j} + P'_2 \frac{\partial x'^2}{\partial x^j} + P'_3 \frac{\partial x'^3}{\partial x^j}.$$

(95)

In (95), the index is $j = 1, 2, 3$, and the components $P'_1$, $P'_2$, $P'_3$ are equal to zero according to (92). From (92) and (95) for the energy and momentum components in the reference frame $K$ we obtain the following

$$E = E_0 \frac{\partial x'^0}{\partial x^0}, \qquad \mathbf{P} = \left(-\frac{E_0}{c}\frac{\partial x'^0}{\partial x^1}, -\frac{E_0}{c}\frac{\partial x'^0}{\partial x^2}, -\frac{E_0}{c}\frac{\partial x'^0}{\partial x^3}\right). \tag{96}$$

Let us define the inertial mass $\mathcal{M}$ of a physical system as a scalar factor by which the four-velocity of the center of momentum must be multiplied in order to obtain the four-momentum in the reference frames $K$ and $K'$:

$$P_\mu = \mathcal{M} u_\mu, \qquad\qquad P'_\nu = \mathcal{M} u'_\nu. \qquad (97)$$

From (91-92) and (96-97) it follows:

$$u'_\nu = (u'_0, 0) = \left(\frac{E_0}{\mathcal{M} c}, 0\right), \qquad E_0 = \mathcal{M} c u'_0.$$

$$E = E_0 \frac{\partial x'^0}{\partial x^0} = \mathcal{M} c u'_0 \frac{\partial x'^0}{\partial x^0}, \qquad \mathbf{P} = \left(-\mathcal{M} u'_0 \frac{\partial x'^0}{\partial x^1}, -\mathcal{M} u'_0 \frac{\partial x'^0}{\partial x^2}, -\mathcal{M} u'_0 \frac{\partial x'^0}{\partial x^3}\right).$$
$$(98)$$

According to (98), the rest energy $E_0$ in the reference frame $K'$ of the center of momentum depends on the time component $u'_0$ of the four-velocity of the center of momentum. To calculate energy $E$ and momentum $\mathbf{P}$ in an arbitrary reference frame $K$, it is also necessary to know the time components of the Jacobian $\frac{\partial x'^0}{\partial x^\mu}$ (94).

In the special theory of relativity, relations (98) are simplified. Let the reference frame $K'$ move along the axis $OX$ of the reference frame $K$ with a constant speed $V$. At the initial moment of time, the center of momentum of the physical system, associated with the origin of the coordinate system $K'$, is at the origin of the coordinate system $K$. In this case, we have $u'_0 = c$, and the Jacobian $S^\nu{}_\mu$ corresponds to the Lorentz transformations and has the following form:

$$S^\nu{}_\mu = \frac{\partial x'^\nu}{\partial x^\mu} = \begin{pmatrix} \frac{1}{\sqrt{1-V^2/c^2}} & -\frac{V}{c\sqrt{1-V^2/c^2}} & 0 & 0 \\ -\frac{V}{c\sqrt{1-V^2/c^2}} & \frac{1}{\sqrt{1-V^2/c^2}} & 0 & 0 \\ 0 & 0 & 1 & 0 \\ 0 & 0 & 0 & 1 \end{pmatrix}. \qquad (99)$$

Taking (99) into account, standard expressions for energy and momentum in the special theory of relativity follow from (98):

$$u'_\nu = (c, 0), \qquad E_0 = \mathcal{M} c^2.$$

$$E = \frac{\mathcal{M} c^2}{\sqrt{1 - V^2/c^2}}, \qquad \mathbf{P} = \left( \frac{\mathcal{M} V}{\sqrt{1 - V^2/c^2}}, 0, 0 \right). \qquad (100)$$

In (14), the volume density $\mathcal{P}_\mu = -\dfrac{\partial \mathcal{L}}{\partial u^\mu}$ of the generalized four-momentum was determined. We can integrate this density over the proper volumes of all the particles and find the generalized four-momentum of the system:

$$p_\mu = \int_{V_0} \mathcal{P}_\mu dV_0 = \frac{1}{c} \int_V \mathcal{P}_\mu u^0 \sqrt{-g}\, dx^1 dx^2 dx^3 = (p_0, -\mathbf{p}) = \left( \frac{E_p}{c}, -\mathbf{p} \right). \qquad (101)$$

By virtue of its construction, the generalized four-momentum $p_\mu$ (101) is a four-vector, and the energy $E_p$ is related to the particle energy expressed through scalar field potentials.

Let us substitute the Lagrangian density (5) into the expression $\mathcal{P}_\mu = -\dfrac{\partial \mathcal{L}}{\partial u^\mu}$ and again assume that the four-potentials and the field tensors do not depend directly on the four-velocity $u^\mu$ of any given matter element. Then relation (19) holds true for $\mathcal{P}_\mu$, and the quantities $E_p$ and $\mathbf{p}$, which are part of the generalized four-momentum, follow from (101):

$$E_p = \int_V \mathcal{P}_0 u^0 \sqrt{-g}\, dx^1 dx^2 dx^3 = \int_V \left( \rho_{0q} A_0 + \rho_0 D_0 + \rho_0 U_0 + \rho_0 \pi_0 \right) u^0 \sqrt{-g}\, dx^1 dx^2 dx^3 =$$

$$= \frac{1}{c} \int_V \left( \rho_{0q} \varphi + \rho_0 \psi + \rho_0 \vartheta + \rho_0 \wp \right) u^0 \sqrt{-g}\, dx^1 dx^2 dx^3.$$

$$\mathbf{p} = \frac{1}{c} \int_V \left( \rho_{0q} \mathbf{A} + \rho_0 \mathbf{D} + \rho_0 \mathbf{U} + \rho_0 \boldsymbol{\pi} \right) u^0 \sqrt{-g}\, dx^1 dx^2 dx^3. \qquad (102)$$

As shown in (102), the energy $E_p$ and the momentum $\mathbf{p}$ of the system's particles are part of the system's energy $E$ (77) and momentum $\mathbf{P}$ (90), respectively. In addition, in case when the four-potentials and the field tensors do not depend directly on the particles' velocities, $\mathbf{p}$ and $\mathbf{P}$ coincide with each other.

### 3.9. Angular momentum

To determine the angular momentum and its conservation law, the property of space isotropy is used when, in the case of simultaneous rotation of all the particles of a closed system about a certain axis by a constant small angular vector $\delta\boldsymbol{\varphi}$, the state of the system does not change [19]. This means that the variation $\delta S$ of the action function and the Lagrangian variation $\delta L$ must be equal to zero. While rotating, the particles' coordinates and their velocities change, and for their variations we can write the following:

$$\delta \mathbf{r}_n = [\delta\boldsymbol{\varphi} \times \mathbf{r}_n], \qquad \delta \mathbf{v}_n = \delta \frac{d\mathbf{r}_n}{dt} = \frac{d(\delta \mathbf{r}_n)}{dt} = [\delta\boldsymbol{\varphi} \times \mathbf{v}_n]. \tag{103}$$

For the Lagrangian variation after permutation of vectors in mixed products, in view of (103) we obtain the following:

$$\delta L = \sum_{n=1}^{N} \frac{\partial L}{\partial \mathbf{r}_n} \delta \mathbf{r}_n + \sum_{n=1}^{N} \frac{\partial L}{\partial \mathbf{v}_n} \delta \mathbf{v}_n = \sum_{n=1}^{N} \frac{\partial L}{\partial \mathbf{r}_n}[\delta\boldsymbol{\varphi} \times \mathbf{r}_n] + \sum_{n=1}^{N} \frac{\partial L}{\partial \mathbf{v}_n}[\delta\boldsymbol{\varphi} \times \mathbf{v}_n] =$$
$$= \delta\boldsymbol{\varphi} \sum_{n=1}^{N} \left[ \mathbf{r}_n \times \frac{\partial L}{\partial \mathbf{r}_n} \right] + \delta\boldsymbol{\varphi} \sum_{n=1}^{N} \left[ \mathbf{v}_n \times \frac{\partial L}{\partial \mathbf{v}_n} \right] = 0. \tag{104}$$

We take into account (72) in (104):

$$\delta L = \delta\boldsymbol{\varphi} \sum_{n=1}^{N} \left[ \mathbf{r}_n \times \frac{d}{dt}\left( \frac{\partial L}{\partial \mathbf{v}_n} \right) \right] + \delta\boldsymbol{\varphi} \sum_{n=1}^{N} \left[ \mathbf{v}_n \times \frac{\partial L}{\partial \mathbf{v}_n} \right] = \delta\boldsymbol{\varphi} \frac{d}{dt} \sum_{n=1}^{N} \left[ \mathbf{r}_n \times \frac{\partial L}{\partial \mathbf{v}_n} \right] = 0. \tag{105}$$

Since $\delta\boldsymbol{\varphi} \neq 0$, then from (105) it follows that in a closed system of particles the angular momentum vector is conserved, which in view of (88) equal to:

$$\mathbf{M} = \sum_{n=1}^{N} \left[ \mathbf{r}_n \times \frac{\partial L}{\partial \mathbf{v}_n} \right] = \sum_{n=1}^{N} [\mathbf{r}_n \times \mathbf{P}_n]. \tag{106}$$

The quantity $\mathbf{P}_n$ is the three-dimensional momentum of one volume element of the system associated with a particle with number $n$.

We substitute into (106) the Lagrangian, which consists of two parts $L = L_p + L_f = \int_V (\mathcal{L}_p + \mathcal{L}_f)\sqrt{-g}\, dx^1 dx^2 dx^3$. In addition, for $L_p = c\int_V \frac{\mathcal{L}_p}{u^0} dV_0$, we can proceed from summation in (106) to integration over the volume:

$$\mathbf{M} = \int_V \left[\mathbf{r} \times \frac{\partial}{\partial \mathbf{v}}\left(\frac{\mathcal{L}_p}{u^0}\right)\right] u^0 \sqrt{-g}\, dx^1 dx^2 dx^3 + \sum_{n=1}^{N}\left[\mathbf{r}_n \times \frac{\partial L_f}{\partial \mathbf{v}_n}\right]. \qquad (107)$$

We can also substitute $\mathcal{L}_p$ from (80) into (107):

$$\mathbf{M} = \frac{1}{c}\int_V \mathbf{r} \times \left[\begin{array}{l} \rho_{0q}\mathbf{A} + \rho_0 \mathbf{D} + \rho_0 \mathbf{U} + \rho_0 \mathbf{\Pi} - \dfrac{\partial}{\partial \mathbf{v}}\left(\rho_{0q}\varphi + \rho_0 \psi + \rho_0 \vartheta + \rho_0 \wp\right) + \\ + \left[\dfrac{\partial}{\partial \mathbf{v}}\left(\rho_{0q}\mathbf{A} + \rho_0 \mathbf{D} + \rho_0 \mathbf{U} + \rho_0 \mathbf{\Pi}\right)\right] \cdot \mathbf{v} \end{array}\right] u^0 \sqrt{-g}\, dx^1 dx^2 dx^3 +$$

$$+ \sum_{n=1}^{N}\left[\mathbf{r}_n \times \frac{\partial L_f}{\partial \mathbf{v}_n}\right].$$

(108)

In (108), we can see that the main part of the angular momentum of the system with continuous matter distribution is created by the vector potentials of all the fields in the matter. The derivatives $\dfrac{\partial}{\partial \mathbf{v}}$ of the fields' potentials also contribute to the results. In (108) there is also an addition that depends on the derivatives of the Lagrangian $L_f$ with respect to the velocities of the system's particles when the field tensor invariants depend on these velocities.

It should be noted that the angular momentum $\mathbf{M}$ is a pseudovector. This is because if we replace the right-hand spatial coordinate system with the left-hand coordinate system, $\mathbf{M}$ will change its sign.

### 3.10. Angular momentum pseudotensor of system of particles

In (91), we defined the four-momentum $P_\mu$, with the help of which we will now define the four-dimensional angular momentum pseudotensor. For the system of $N$ particles, this pseudotensor with covariant indices is calculated by the following formula:

$$M_{\mu\nu} = \sum_{n=1}^{N}\left(x_\mu P_\nu - x_\nu P_\mu\right)_n = \sum_{n=1}^{N}\left(x_\mu \times P_\nu\right)_n. \tag{109}$$

In (109) all the quantities in brackets refer to one particle with the current number $n$, and the pseudotensor for the entire system is found by summing on all the particles. In brackets, we find the vector product of two quantities – the position vector and four-momentum taken with covariant indices. Since the position vector $x_\mu$ is not a four-vector, this justifies the name of a pseudotensor for $M_{\mu\nu}$.

We find the individual components of the pseudotensor (109) within the framework of the special theory of relativity, for which we use the expressions for the position vector and four-momentum (91) for each particle in the Cartesian coordinate system:

$$\left(x_\mu\right)_n = (ct, -\mathbf{r}_n) = (ct, -x_n, -y_n, -z_n), \quad \left(P_\nu\right)_n = \left(\frac{E_n}{c}, -\mathbf{P}_n\right) = \left(\frac{E_n}{c}, -P_{xn}, -P_{yn}, -P_{zn}\right).$$

$$M_{01} = -M_{10} = \sum_{n=1}^{N}\left(x_0 P_1 - x_1 P_0\right)_n = -\sum_{n=1}^{N}\left(ct P_{xn} - \frac{1}{c} x_n E_n\right),$$

$$M_{02} = -M_{20} = \sum_{n=1}^{N}\left(x_0 P_2 - x_2 P_0\right)_n = -\sum_{n=1}^{N}\left(ct P_{yn} - \frac{1}{c} y_n E_n\right),$$

$$M_{03} = -M_{30} = \sum_{n=1}^{N}\left(x_0 P_3 - x_3 P_0\right)_n = -\sum_{n=1}^{N}\left(ct P_{zn} - \frac{1}{c} z_n E_n\right),$$

$$M_{12} = -M_{21} = \sum_{n=1}^{N}\left(x_1 P_2 - x_2 P_1\right)_n = \sum_{n=1}^{N}\left(x_n P_{yn} - y_n P_{xn}\right) = \sum_{n=1}^{N}\left[\mathbf{r}_n \times \mathbf{P}_n\right]_z,$$

$$M_{13} = -M_{31} = \sum_{n=1}^{N}\left(x_1 P_3 - x_3 P_1\right)_n = \sum_{n=1}^{N}\left(x_n P_{zn} - z_n P_{xn}\right) = -\sum_{n=1}^{N}\left[\mathbf{r}_n \times \mathbf{P}_n\right]_y,$$

$$M_{23} = -M_{32} = \sum_{n=1}^{N}\left(x_2 P_3 - x_3 P_2\right)_n = \sum_{n=1}^{N}\left(y_n P_{zn} - z_n P_{yn}\right) = \sum_{n=1}^{N}\left[\mathbf{r}_n \times \mathbf{P}_n\right]_x.$$

$$\tag{110}$$

Since the pseudotensor is antisymmetric, its components $M_{00}$, $M_{11}$, $M_{22}$ and $M_{33}$ are equal to zero. We can see from (110) that the pseudotensor components are two three-dimensional vectors. One of these vectors consists of the components $M_{01}$, $M_{02}$ and $M_{03}$, and is written as follows:

$$\mathbf{C} = -\sum_{n=1}^{N}\left(ct\mathbf{P}_n - \frac{1}{c}E_n \mathbf{r}_n\right). \tag{111}$$

The other vector consists of the components $M_{23}$, $M_{31}$, $M_{12}$ and represents the system's angular momentum in the form of (106):

$$\mathbf{M} = \sum_{n=1}^{N}\left[\mathbf{r}_n \times \mathbf{P}_n\right]. \tag{112}$$

Taking into account (111-112), the angular momentum pseudotensor of the system of particle takes the following form:

$$M_{\mu\nu} = \begin{pmatrix} 0 & C_x & C_y & C_z \\ -C_x & 0 & M_z & -M_y \\ -C_y & -M_z & 0 & M_x \\ -C_z & M_y & -M_x & 0 \end{pmatrix}. \tag{113}$$

**3.11. Angular momentum pseudotensor in the case of continuous distribution of matter**

To turn to the continuous medium approximation in (109), $P_\mu$ should be replaced by $dP_\mu$ as an equivalent of the four-momentum for one volume element, and the sum over the particles should be replaced by the volume integral:

$$M_{\mu\nu} = \int\left(x_\mu \, dP_\nu - x_\nu \, dP_\mu\right) = \int\left(x_\mu \times dP_\nu\right). \tag{114}$$

If in [2] the angular momentum pseudotensor is defined in the form $M^{\mu\nu}$, that is, with contravariant components, then in (114) the primary expression $M_{\mu\nu}$ is with covariant indices. This is because the definition of a four-momentum is its expression $P_\mu$ with a covariant index.

The components of the pseudotensor $M_{\mu\nu}$ in (114) in Cartesian coordinates are as follows:

$$M_{01} = -M_{10} = \int (x_0\, dP_1 - x_1\, dP_0) = -\int \left( ct\, dP_x - \frac{1}{c} x\, dE \right),$$

$$M_{02} = -M_{20} = \int (x_0\, dP_2 - x_2\, dP_0) = -\int \left( ct\, dP_y - \frac{1}{c} y\, dE \right),$$

$$M_{03} = -M_{30} = \int (x_0\, dP_3 - x_3\, dP_0) = -\int \left( ct\, dP_z - \frac{1}{c} z\, dE \right),$$

$$M_{12} = -M_{21} = \int (x_1\, dP_2 - x_2\, dP_1) = \int (x\, dP_y - y\, dP_x) = \int [\mathbf{r} \times d\mathbf{P}]_z,$$

$$M_{13} = -M_{31} = \int (x_1\, dP_3 - x_3\, dP_1) = \int (x\, dP_z - z\, dP_x) = -\int [\mathbf{r} \times d\mathbf{P}]_y,$$

$$M_{23} = -M_{32} = \int (x_2\, dP_3 - x_3\, dP_2) = \int (y\, dP_z - z\, dP_y) = \int [\mathbf{r} \times d\mathbf{P}]_x. \qquad (115)$$

In (115) the pseudotensor components $M_{01}$, $M_{02}$ and $M_{03}$ form a three-dimensional vector:

$$\mathbf{C} = -\int \left( ct\, d\mathbf{P} - \frac{1}{c} \mathbf{r}\, dE \right). \qquad (116)$$

Moreover, the pseudotensor components $M_{23}$, $M_{31}$ and $M_{12}$ form another three-dimensional vector, which is the system's angular momentum in integral form:

$$\mathbf{M} = \int [\mathbf{r} \times d\mathbf{P}]. \qquad (117)$$

The components of vectors $\mathbf{C}$ (116) and $\mathbf{M}$ (117) form the components of the angular momentum pseudotensor $M_{\mu\nu}$ for a continuous medium in the same form as in (113) for the case of a system of particle.

**3.12. Angular momentum pseudotensor of a body taking into account its fields**

In the general case, the components of the angular momentum pseudotensor must be found by summing the components in the continuously distributed matter of the body, with the corresponding components associated with fields outside the body. For the volume occupied by matter, formulas (116-117) should be used, and for the volume outside matter, where there are only fields, formulas (111-112) should be used.

We will proceed from the expression (89) for the momentum of the system taking into account the Lagrangian $L = L_p + L_f = \int_V (\mathcal{L}_p + \mathcal{L}_f) \sqrt{-g}\, dx^1 dx^2 dx^3$, in which the Lagrangian density $\mathcal{L} = \mathcal{L}_p + \mathcal{L}_f$ consists of two parts. For the volume occupied by matter, the momentum differential is equal to:

$$d\mathbf{P}_m = \frac{\partial}{\partial \mathbf{v}}\left(\frac{\mathcal{L}_p}{u^0}\right) u^0 \sqrt{-g}\, dx^1 dx^2 dx^3 = c \frac{\partial}{\partial \mathbf{v}}\left(\frac{\mathcal{L}_p}{u^0}\right) dV_0. \qquad (118)$$

The fields also contribute to the momentum of one volume element through $\mathcal{L}_f$, taking into account the definition $L_f = \int_{V_n} \mathcal{L}_f \sqrt{-g}\, dx^1 dx^2 dx^3$:

$$\mathbf{P}_{nf} = \frac{\partial L_f}{\partial \mathbf{v}_n} = \frac{\partial}{\partial \mathbf{v}_n}\left(\int_{V_n} \mathcal{L}_f \sqrt{-g}\, dx^1 dx^2 dx^3\right). \qquad (119)$$

Substituting $d\mathbf{P}_m$ (118) into (117), and $\mathbf{P}_{nf}$ (119) into (112), and summing up the results, we arrive at the angular momentum of the system in the form (107):

$$\mathbf{M} = \int_V \left[\mathbf{r} \times \frac{\partial}{\partial \mathbf{v}}\left(\frac{\mathcal{L}_p}{u^0}\right)\right] u^0 \sqrt{-g}\, dx^1 dx^2 dx^3 + \sum_{n=1}^{N}\left[\mathbf{r}_n \times \frac{\partial L_f}{\partial \mathbf{v}_n}\right]. \qquad (120)$$

In (120), one part of the angular momentum $\mathbf{M}$ is expressed through an integral, and the other part is expressed through the sum over all particles of the system.

The differential of the integral part of the system's energy and the energy of one volume element that depends on the derivative $\dfrac{\partial L_f}{\partial \mathbf{v}_n}$, according to (76), are equal:

$$dE = \left[ \mathbf{v} \cdot \frac{\partial}{\partial \mathbf{v}} \left( \frac{\mathcal{L}_p}{u^0} \right) - \frac{\mathcal{L}_p}{u^0} \right] u^0 \sqrt{-g}\, dx^1 dx^2 dx^3 - \mathcal{L}_f \sqrt{-g}\, dx^1 dx^2 dx^3. \tag{121}$$

$$E_{nf} = \mathbf{v}_n \cdot \frac{\partial L_f}{\partial \mathbf{v}_n}. \tag{122}$$

Substituting $d\mathbf{P}_m$ (118) and $dE$ (121) into (116), we find the vector $\mathbf{C}_1$ related to the volume occupied by matter of the body:

$$\begin{aligned} \mathbf{C}_1 &= -\int \left( ct\, d\mathbf{P} - \frac{1}{c} \mathbf{r}\, dE \right) = \\ &= -\int \left\{ ct \frac{\partial}{\partial \mathbf{v}} \left( \frac{\mathcal{L}_p}{u^0} \right) u^0 - \frac{1}{c} \mathbf{r} \left[ \mathbf{v} \cdot \frac{\partial}{\partial \mathbf{v}} \left( \frac{\mathcal{L}_p}{u^0} \right) u^0 - \mathcal{L}_p - \mathcal{L}_f \right] \right\} \sqrt{-g}\, dx^1 dx^2 dx^3. \end{aligned} \tag{123}$$

Let us now substitute $\mathbf{P}_{nf}$ (119) and $E_{nf}$ (122) into (111), and find the vector $\mathbf{C}_2$ associated with the fields:

$$\mathbf{C}_2 = -\sum_{n=1}^{N} \left[ ct \frac{\partial L_f}{\partial \mathbf{v}_n} - \frac{1}{c} \left( \mathbf{v}_n \cdot \frac{\partial L_f}{\partial \mathbf{v}_n} \right) \mathbf{r}_n \right]. \tag{124}$$

Summing up $\mathbf{C}_1$ (123) and $\mathbf{C}_2$ (124), we find the so-called time-varying dynamic mass moment:

$$\begin{aligned} \mathbf{C} = \mathbf{C}_1 + \mathbf{C}_2 &= -\int \left\{ ct \frac{\partial}{\partial \mathbf{v}} \left( \frac{\mathcal{L}_p}{u^0} \right) u^0 - \frac{1}{c} \mathbf{r} \left[ \mathbf{v} \cdot \frac{\partial}{\partial \mathbf{v}} \left( \frac{\mathcal{L}_p}{u^0} \right) u^0 - \mathcal{L}_p - \mathcal{L}_f \right] \right\} \sqrt{-g}\, dx^1 dx^2 dx^3 - \\ &- \sum_{n=1}^{N} \left[ ct \frac{\partial L_f}{\partial \mathbf{v}_n} - \frac{1}{c} \left( \mathbf{v}_n \cdot \frac{\partial L_f}{\partial \mathbf{v}_n} \right) \mathbf{r}_n \right]. \end{aligned}$$

(125)

In (125) it is convenient to use the Lagrangian density component $\mathcal{L}_p$ in the form (80), $L_f = \int_{V_n} \mathcal{L}_f \sqrt{-g}\, dx^1 dx^2 dx^3$, where $\mathcal{L}_f$ is given in (27). In this case, we should take into account the energy calibration condition in the form $R = 2\Lambda$, which follows from the first relation in (40).

In the mass moment **C** (125), we select terms containing the radius-vectors **r** and $\mathbf{r}_n$, and with their help we determine the radius-vector of the center of momentum of the system under consideration:

$$\mathbf{R} = \frac{1}{E}\left\{\int \mathbf{r}\left[\mathbf{v}\cdot\frac{\partial}{\partial \mathbf{v}}\left(\frac{\mathcal{L}_p}{u^0}\right)u^0 - \mathcal{L}_p - \mathcal{L}_f\right]\sqrt{-g}\,dx^1 dx^2 dx^3 + \sum_{n=1}^{N}\left(\mathbf{v}_n\cdot\frac{\partial L_f}{\partial \mathbf{v}_n}\right)\mathbf{r}_n\right\} \quad (126)$$

Within the framework of the special theory of relativity, we can assume that the center of momentum of a closed system moves at a certain constant velocity **V**. In this case, the momentum **P** holds the relation $\mathbf{P} = \frac{E}{c^2}\mathbf{V}$ where the energy $E$ of the system is proportional to the Lorentz factor $\gamma = \frac{1}{\sqrt{1 - V^2/c^2}}$, the inertial mass of the system and the square of the speed of light. Taking this into account, as well as the expressions **P** in (89) and **R** in (126), for the mass moment (125) we have:

$$\mathbf{C} = -ct\mathbf{P} + \frac{E}{c}\mathbf{R} = \frac{E}{c}(\mathbf{R} - \mathbf{V}t). \quad (127)$$

In a closed system, the pseudotensor $M_{\mu\nu}$ is conserved, and its components are constant values. This approach implies conservation of the angular momentum, $\mathbf{M} = const$, as well as conservation of the mass moment $\mathbf{C} = \frac{E}{c}(\mathbf{R} - \mathbf{V}t) = const$ and of the energy $E = const$ of the system, In this case, we can introduce a constant radius-vector $\mathbf{R}_0 = \frac{c\mathbf{C}}{E}$, expressing the position of the system's center of momentum at $t = 0$. Then, the equation of rectilinear motion of the center of momentum of the closed system follows from (127): $\mathbf{R} = \mathbf{R}_0 + \mathbf{V}t$.

## 4. Conclusion

The main purpose of using the Lagrangian formalism is derivation from the principle of least action of the field equations, equation for the metric and equations of motion for each particular Lagrangian. A more in-depth approach provides, in general form, formulas relating the Lagrangian density to the corresponding equation or integral of motion.

As an example we can mention the general Euler–Lagrange equation (2), equation of motion (13), equation for the metric (31), formula for the stress-energy tensor (42), Euler–Lagrange equation for vector fields by the example of the electromagnetic field (51); formulas for the energy (75-76), Hamiltonian (78), Hamiltonian equations (83-84); momentum (88-89); generalized four-momentum (101); angular momentum (106-107); four-dimensional angular momentum pseudotensor (113); mass moment (125); radius-vector of the center of momentum (126).

When presenting the results, we paid special attention to derivation of covariantly written formulas suitable for use in continuously distributed matter in curved spacetime. Some of these formulas have been simplified and written as is usually accepted in the special theory of relativity.

In (15) we presented the equation of motion of a physical system, expressed in terms of derivatives of field potentials. This led us to the concepts of generalized four-momentum density (14) and generalized four-force density (15). Next, we presented the equation for the metric, the equations for the fields, and the equation of motion written using field tensors. The following sections show how energy, Hamiltonian, momentum, four-momentum, angular momentum, and the angular momentum pseudotensor are defined using the Lagrangian formalism. Thus, for vector fields it turns out to be possible to derive the main physical quantities in a covariant way. These quantities are necessary to describe the equations of motion and evolution of physical systems containing continuously distributed matter and basic acting fields, such as gravitational and electromagnetic fields, acceleration field and pressure field.

**Declaration of interests**

The authors declare that they have no known competing financial interests or personal relationships that could have appeared to influence the work reported in this paper.